\documentclass[sigconf,nonacm]{acmart}

\usepackage{enumitem}
\usepackage{graphbox}
\usepackage{listings}
\usepackage{subcaption}
\usepackage{makecell}
\usepackage{balance}
\usepackage{amsmath}
\usepackage{xcolor}
\usepackage{algcompatible}
\usepackage{soul}
\usepackage{xspace}
\usepackage{hyperref}
\usepackage{diagbox}
\usepackage[utf8]{inputenc}
\hypersetup{pdfencoding=unicode}
\inputencoding{utf8}
\makeatother

\usepackage[UKenglish]{babel}
\usepackage{colorprofiles}
\hypersetup{pdfstartview=}
\usepackage{cleveref}
\usepackage{booktabs}
\usepackage{array}
\usepackage[labelformat=simple,subrefformat=parens]{subcaption}

\AtBeginDocument{%
  }

\setcopyright{acmlicensed}
\copyrightyear{2018}
\acmYear{2018}
\acmDOI{XXXXXXX.XXXXXXX}
\acmConference[Conference acronym 'XX]{Make sure to enter the correct
  conference title from your rights confirmation email}{June 03--05,
  2018}{Woodstock, NY}
\acmISBN{978-1-4503-XXXX-X/2018/06}

\begin{document}

\definecolor{caribbeangreen}{rgb}{0.0, 0.8, 0.6}
\newcommand{\hcha}[1]{\textbf{\color{violet}[hk] #1}}
\newcommand{\yxy}[1]{{\color{red}[yxy] #1}}
\newcommand{\aditya}[1]{\textbf{\color{orange}[aditya] #1}}
\newcommand{\rev}[1]{\textbf{\color{blue}#1}}

\newcommand{\nowait}{NO\allowbreak\textunderscore\allowbreak WAIT\xspace}
\newcommand{\waitdie}{WAIT\allowbreak\textunderscore\allowbreak DIE\xspace}
\newcommand{\woundwait}{WOUND\allowbreak\textunderscore\allowbreak WAIT\xspace}
\newcommand{\work}{Lotus\xspace}

\title{Two-sided RDMA Striking Back for Disaggregated Memory Databases}

% \author{Paper ID: 411}

\author{Hokeun Cha}
\email{hcha@cs.wisc.edu}
\affiliation{%
 \institution{University of Wisconsin-Madison}
 \city{Madison}
 \state{WI}
 \country{USA}
}

\author{Aditya Akella}
\email{akella@cs.utexas.edu}
\affiliation{%
 \institution{University of Texas at Austin}
 \city{Austin}
 \state{TX}
 \country{USA}
}

\author{Xiangyao Yu}
\email{yxy@cs.wisc.edu}
\affiliation{%
 \institution{University of Wisconsin-Madison}
 \city{Madison}
 \state{WI}
 \country{USA}
}



% \keywords{Do, Not, Use, This, Code, Put, the, Correct, Terms, for,


\begin{abstract}

RDMA has enabled high-speed data access and low-latency communication in disaggregated memory databases.
While various optimization techniques have been proposed to accelerate transactions with RDMA in this setting,
two-sided RDMA has been largely underexplored in favor of one-sided RDMA due to its remote CPU involvement.
% \yxy{<-- our argument here is true only in the scope of memory disaggregation (I think this is true?); to ensure readers don't misunderstand and start to feel skeptical, maybe somehow mention/highlight disaggregation in this sentence. or maybe merge this with the 1st sentence..} \hcha{You're right. Added mentioning about setting. }
% two-sided RDMA has been largely overlooked in favor of one-sided RDMA due to its remote CPU involvement.
However, the heavy use of one-sided RDMA introduces fundamental limitations.
Its limited APIs cannot express complex system functions such as starvation prevention, priority-based scheduling, and preemption, which are all critical functions in concurrency control protocols.
% Moreover, the limited APIs require multiple network round-trips to process indexing requests, causing network amplification.
Moreover, indexing requires multiple network round-trips, causing network amplification.

In this work, we revisit the long-standing debate between one-sided RDMA and two-sided RDMA in the context of disaggregated memory databases.
We present \work, which addresses the conventional limitation of two-sided RDMA---CPU bottlenecks in memory servers---by leveraging the rich functionality of two-sided RDMA with two key optimization techniques: (1) lightweight caching and (2) efficient batching.
\work demonstrates that limited CPU resources in memory servers, when intelligently utilized, can transform a perceived weakness into a significant advantage.
Our experimental study shows that \work achieves up to 8.2$\times$ higher throughput and 42.9$\times$ lower p999 tail latency than state-of-the-art one-sided RDMA-based approaches in YCSB benchmark.

\end{abstract}

\maketitle

\section{Introduction}
\label{sec:intro}

Remote Direct Memory Access (RDMA) has emerged as a cornerstone technology for high-performance distributed database systems over the past decade, 
offering unprecedented network performance with low latency and high bandwidth.
The systems research community has extensively studied RDMA for its ability to dramatically improve the performance of distributed OLTP databases,
leading to significant advances in indexing~\cite{namtree, sherman, race, rolex}, concurrency control~\cite{drtm, drtmh, pilaf, farm, farm2}, and data replication~\cite{tebis, tailwind, active_memory}.

The RDMA research has been marked by a fundamental debate: whether to use one-sided RDMA~\cite{drtm, pilaf, farm, farm2} or two-sided RDMA~\cite{herd, fasst, eRPC, scaleRPC} in network communication.
This debate has shaped the evolution of RDMA-based database systems,
with each approach offering distinct trade-offs that have influenced design decisions across the research community.

The salient feature of one-sided RDMA, i.e., direct access to remote memory without remote CPU involvement, has opened up a new paradigm in system designs.
In contrast, two-sided RDMA has enabled smooth transition for existing systems via its socket-like APIs.
However, two-sided RDMA has been largely underexplored in disaggregated memory, where compute and memory resources are decoupled.
% While both have merits, two-sided RDMA has been largely underexplored in disaggregated memory, where compute and memory resources are decoupled.
% While both have their merits, two-sided RDMA has been largely overlooked in disaggregated memory systems, where compute and memory resources are separated.
In such settings, where memory servers typically have limited computing resources compared to compute servers, 
% In such systems, where computing resources in memory servers are limited compared to those in compute servers, 
two-sided RDMA systems were considered suboptimal for handling a high volume of requests from compute servers.
% \yxy{(please check) suboptimal to} handle the large number of requests from compute servers.

% disaggregated memory systems, where compute and memory resources are separated.
% However, its APIs are limited to simple memory operations, which cannot express complex system functions efficiently.
% Examples include starvation prevention, priority-based scheduling, and preemption---all of them are critical functions in concurrency control protocols.
% Also, the limited APIs often require multiple network round-trips to complete a single request. e.g., indexing, increasing latency and network traffic.
% % To address these limitations prior work has proposed various techniques, such as caching~\cite{sherman, deft, smart} and compute-level partitioning~\cite{dex}.

% On the other hand, two-sided RDMA has enabled smooth transition for existing applications due to its socket-like APIs, preserving rich functional features.
% However, it has been overlooked in disaggregated memory systems, where compute and memory resources are separated, because two-sided RDMA requires CPU involvement on both sides in compute servers and memory servers.
% In such systems, where computing resources in memory servers are limited compared to those in compute servers, two-sided RDMA systems in memory servers cannot handle a large number of requests from compute servers.

In this work, we revisit the legacy one-sided versus two-sided RDMA debate in the context of memory disaggregation and argue that the common wisdom favoring one-sided RDMA needs careful review.
Our analysis reveals that two-sided RDMA is a compelling alternative that can overcome its limitations and outperform one-sided RDMA in B+-tree indexing and two-phase locking concurrency control.
% Our analysis reveals that two-sided RDMA is a compelling alternative that can overcome its conventional limitations and even outperform one-sided RDMA in disaggregated memory databases.
In particular, two-sided RDMA in memory-disaggregated OLTP databases can leverage three key features:
% (1) \textit{rich functionality} that can express complex system functions efficiently,
% (2) \textit{caching} to reduce the overhead of remote data processing, and
% (3) \textit{request batching} to minimize the number of network round-trips.
% we show that the optimization techniques for one-sided RDMA, e.g., caching, can be also applied to two-sided RDMA, 
% and demonstrate that, even with limited computing power, two-sided RDMA can outperform one-sided RDMA approaches.

\textbf{(1) Rich Functionality.}
Two-sided RDMA requires remote CPU involvement, which has been considered a major drawback due to limited CPU resources in memory servers.
However, this involvement enables rich functional features that can express complex system functions that are not easily achievable with one-sided RDMA.
% For example, one-sided RDMA makes it extremely difficult to implement starvation prevention, priority-based scheduling, and preemption, all of which are critical functions in concurrency control protocols.
For example, one-sided RDMA makes it extremely difficult to implement starvation prevention, priority-based scheduling, and preemption---all of which are critical functions in concurrency control protocols.
% For example, with one-sided RDMA, it is extremely difficult to implement starvation prevention, priority-based scheduling, and preemption---all of which are critical functions in concurrency control protocols.
Moreover, one-sided RDMA often requires multiple network round-trips for index operations.
Index traversal requires reading the entire node data over the network, and each node access requires at least three RDMA READs to ensure correctness in optimistic synchronization,
i.e., checking a version, reading node data, and validating the version~\cite{rdma_guideline}. 
In contrast, these operations require much fewer network round-trips with two-sided RDMA.
% \yxy{In contrast, these operations require much fewer network round-trips with two-sided RDMA.}

\textbf{(2) Caching.}
To mitigate the network amplification problem in one-sided RDMA indexing, prior work has employed caching in compute servers to reduce remote memory accesses~\cite{sherman, deft, smart, dex}.
While caching has shown effectiveness in one-sided RDMA indexing, it has not been explored in the context of two-sided RDMA indexing.
We make a key observation that these caching techniques can also be applied to two-sided RDMA to avoid unnecessary network round-trips, which can significantly improve performance.
% \yxy{to significantly improve performance}.

% Although caching provides significant performance improvement, it also increases the system complexity due to cache coherence management since multiple compute servers may cache the same data.
% Prior work has proposed various techniques to address this challenge.
% One approach is to cache shortcut hints to data tuples and postpone the cache coherence to later data accesses~\cite{sherman, deft}.
% That is, the system uses a hint to locate the remote data and validates its staleness by checking its metadata.
% Another recent approach is to partition the DBMS in compute layer to avoid cache coherence across compute servers~\cite{dex}.
% Each compute server is assigned to a specific data partition so that the coherence management is only required within a compute server.

% While these techniques have shown effectiveness in one-sided RDMA-based indexing, they have not been explored in the context of two-sided RDMA-based indexing.
% We also apply these two caching techniques to two-sided RDMA-based indexing to avoid unnecessary network round-trips, 
% and show that they are effective in improving the performance of two-sided RDMA-based indexing as well.

\textbf{(3) Batching.}
While two-sided RDMA has been compared with one-sided RDMA in prior work~\cite{herd, fasst, eRPC, scaleRPC}, 
these studies primarily target key-value stores or remote procedure call (RPC) systems that process one request at a time.
However, a database transaction usually consists of multiple data access requests.
Those requests within a transaction can be batched into a network message, which can substantially reduce the number of network round-trips.
% Regardless, some transactions may only have a few requests or may still require multiple round-trips due to data dependencies, which cannot exhibit sufficient batching effects.
Regardless, some transactions may only have a few requests or have data dependency, which cannot exhibit sufficient batching effects.
In such cases, requests from different transactions can be combined into a single network message to further reduce network overhead.
% These batching techniques, i.e., intra-transaction and inter-transaction batching, can substantially alleviate the CPU bottleneck in memory servers.
% \yxy{this paragraph is a bit verbose, especially the intra- and inter-txn batching discussion. Shorten it.}\hcha{Shortened it.}

Motivated by these observations, we design and implement \work, a two-sided RDMA-based system for disaggregated memory OLTP databases that leverages rich functionality, caching, and batching techniques.
Following prior research on RDMA-accelerated OLTP systems~\cite{drtm, drtmh, farm, farm2, pilaf, namdb, sherman, race, dex, deft},
we evaluate the performance of \work against one-sided RDMA schemes in indexing and concurrency control.
In particular, we focus on B+-tree indexing and two-phase locking (2PL), the most widely-used classes of indexing and concurrency control schemes in DBMSs.
Our performance study shows that \work is starvation free and achieves up to 8.2$\times$ higher throughput and 42.9$\times$ lower p999 tail latency compared to state-of-the-art one-sided RDMA schemes in YCSB workloads.
% \yxy{mention that we can achieve starvation freedom?}

In summary, this paper makes the following contributions:\vspace{-.05in}
\begin{itemize}[leftmargin=*]
    \item We revisit the legacy one-sided RDMA versus two-sided RDMA debate in the context of disaggregated memory databases and analyze the advantages and disadvantages of both modes.
    % \yxy{maybe say analyze the advantages and disadvantages of both modes, instead of saying identifying limitations of 1-sided} of one-sided RDMA, i.e., functional deficiency and network amplification.
    \item We develop \work, a scalable two-sided RDMA-based mechanism that efficiently utilizes limited memory server CPU resources by minimizing network overhead through caching and batching.
    \item We comprehensively evaluate \work against one-sided RDMA-based indexing and concurrency control schemes and show that \work outperforms them in both throughput and tail latency.
\end{itemize}

The rest of this paper is organized as follows.
Section~\ref{sec:motivation} presents the background and motivation of the research.
Section~\ref{sec:overview} provides a design overview of \work.
Section~\ref{sec:concurrency_control} describes concurrency control in \work.
We discuss its optimization techniques, caching and batching, in Sections~\ref{sec:cache} and \ref{sec:batch}, respectively.
Section~\ref{sec:evaluation} evaluates \work against the state-of-the-art one-sided RDMA schemes.
Section~\ref{sec:related} reviews related work, and Section~\ref{sec:conclusion} concludes the paper.

\section{Background and Motivation}
\label{sec:motivation}

RDMA has been a crucial component for optimizing distributed database systems due to its high performance and ability to directly access remote memory, particularly in disaggregated memory architectures.
In this section, we review the background of disaggregated memory architectures and RDMA, and discuss the designs and limitations of RDMA-based indexing and concurrency control.  

\begin{figure}[!t]
    \centering
    \begin{minipage}{0.47\textwidth}
        \centering
        \includegraphics[width=\textwidth]{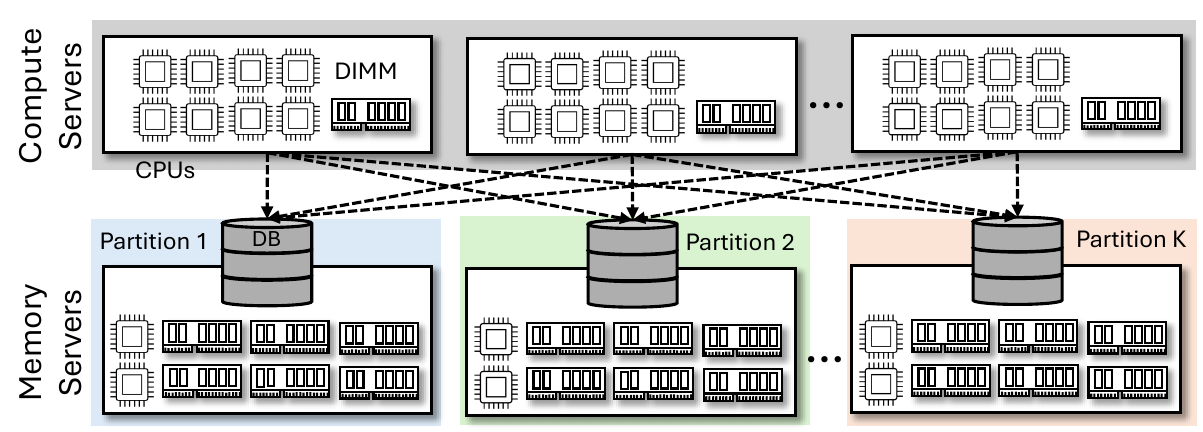}\vspace{-.1in}
        \subcaption{Non-partitioned memory.}
        \label{fig:non_partitioned}
    \end{minipage}
    \hfill
    \centering
    \begin{minipage}{0.47\textwidth}
        \centering
        \includegraphics[width=\textwidth]{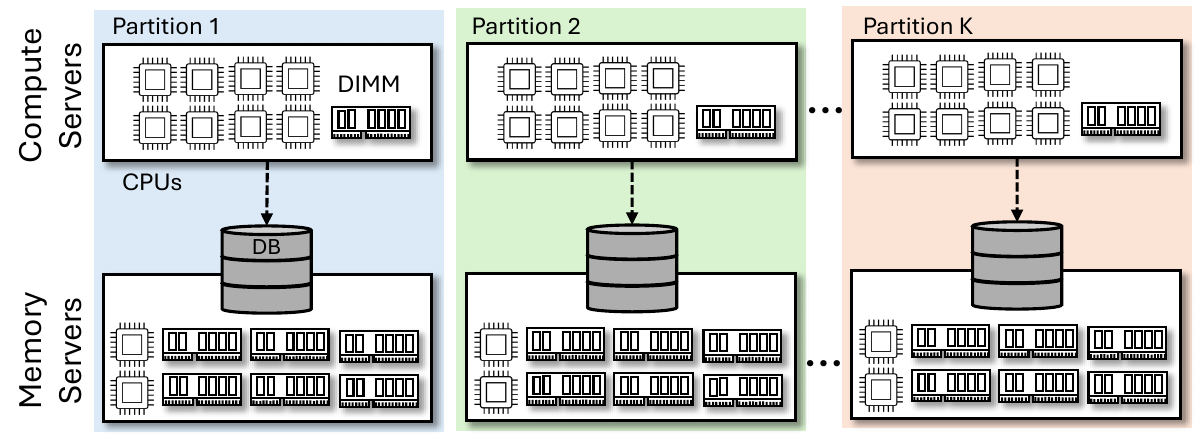}\vspace{-.1in}
        \subcaption{Co-partitioned memory.}\vspace{-.15in}
        % \subcaption{Shared-nothing database.}
        \label{fig:co_partitioned}
    \end{minipage}
    \caption{Memory disaggregation models.}\vspace{-.2in}
    \label{fig:disaggregated_models}
\end{figure}

\subsection{Disaggregated Memory Architecture}
\label{sec:disaggregation}

The emergence of memory disaggregation has led to a redesign of traditional distributed database systems to support important features such as scalability, elasticity, and cost efficiency~\cite{disaggregated_db, aurora, socrates, ballonstasher, namdb, dsmdb}.
Unlike legacy monolithic server architectures, where database components are tightly coupled together, a memory-disaggregated database is decoupled into two layers of servers, \textit{memory servers} and \textit{compute servers}.
A memory server is equipped with a large amount of memory resources and small computing resources.
The large memory space stores data structures and tables for a DBMS.
A compute server has powerful computational resources, while its memory resources are limited.
The two layers of servers can scale independently to strike the best balance of the resources.
% , and they are connected with a high-performance network such as RDMA to mitigate the latency of network communication.
% \yxy{not sure whether the following claim is true or worth mentioning here: In real deployment, the memory and compute servers can be deployed on the same set of physical machines. I feel the claim is incorrect, but it is how we implement it in our testbed. Maybe our implementation is just \textit{simulating} memory disaggregation but not actually achieving it? }
% \hcha{I agree. I removed the last sentence to make it not limited to its implementation. I also changed the names of the system models (shared-memory --> non-partitioned memory, and shared-nothing --> copartitioned memory).}

% \yxy{I feel "shared-memory" and "copartitioned-memory" are probably better names. "shared-nothing" is not a good name since the term has been taken, which is in contradiction to a disaggregation architecture.}
% We name two representative memory disaggregation models, \textit{non-partitioned memory} and \textit{co-partitioned memory}, as shown in Figure~\ref{fig:disaggregated_models}.
Figure~\ref{fig:disaggregated_models} illustrates two representative memory disaggregation models, {\textit{non-partitioned memory}} and {\textit{co-partitioned memory}}. 
In non-partitioned memory, a DBMS is partitioned across memory servers, and each compute server can access any partition in memory servers, providing a globally shared view.
% Non-partitioned memory partitions or shards the DBMS across memory servers, and each compute server can access any partition/shard of the DBMS in memory servers, allowing a globally shared view.
In co-partitioned memory, the DBMS is additionally partitioned in compute servers so that each compute server accesses only its own partition, providing a local view.
% Co-partitioned memory, on the other hand, additionally partitions or shards the DBMS in compute servers so that each compute server only accesses its own partition/shard, providing a local view.
Note that the partitioning functions in the two server layers can be different from each other to efficiently balance the load across servers.
%\yxy{this claim is true only for non-partitioned memory?}\hcha{the claim that the two layers can have different partitioning functions is true only for copartitioned memory. non-partitioned memory does not partition compute servers since every compute server can access any partition of the DBMS in memory servers.}
While transactions, in both memory models, are processed in compute servers by accessing data structures and tables in memory servers,
the difference in the models leads to different behaviors in some database components, e.g., cache coherence and concurrency control.
We discuss the details in Section~\ref{sec:rdma_txn}.

% Transactions, in these architectures, are processed in compute servers by accessing data structures and tables of a DBMS in memory servers.
% The compute servers and memory servers are connected with a high-performance network link such as RDMA to mitigate the latency of remote memory accesses.
% While transactions in these two architectures are both processed in compute servers by accessing data structures and tables in memory servers,
% their behavior is fundamentally different under cross-partition transactions.
% We discuss the details in Section~\ref{sec:rdma_txn}.

% Transactions, in this disaggregated memory architecture, are processed in compute servers by accessing data structures and tables of a DBMS in memory servers. 
% The compute servers and memory servers are connected with a high-performance network such as RDMA to mitigate the latency of remote memory accesses.

\begin{figure*}[!t]
    \centering
    \begin{minipage}{0.48\textwidth}
        \centering
        \includegraphics[width=\textwidth]{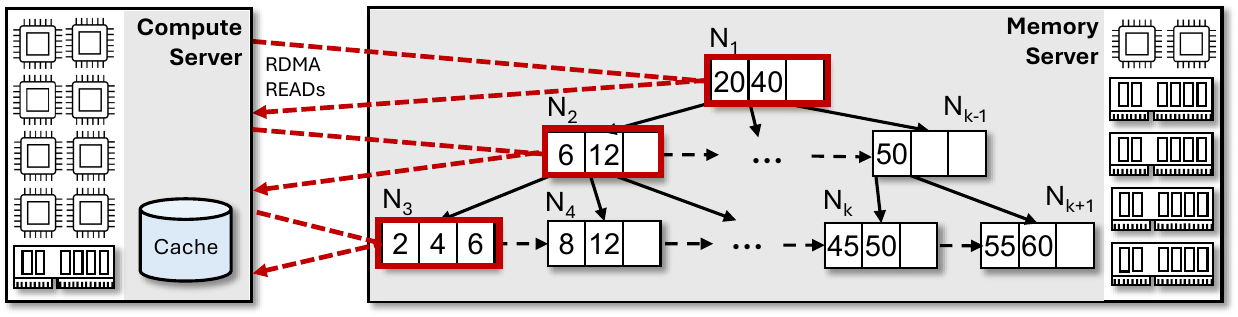}\vspace{-.05in}
        \subcaption{One-sided RDMA indexing.}\vspace{-.15in}
        \label{fig:indexing_scenario_onesided}
    \end{minipage}
    \hfill
    \centering
    \begin{minipage}{0.48\textwidth}
        \centering
        \includegraphics[width=\textwidth]{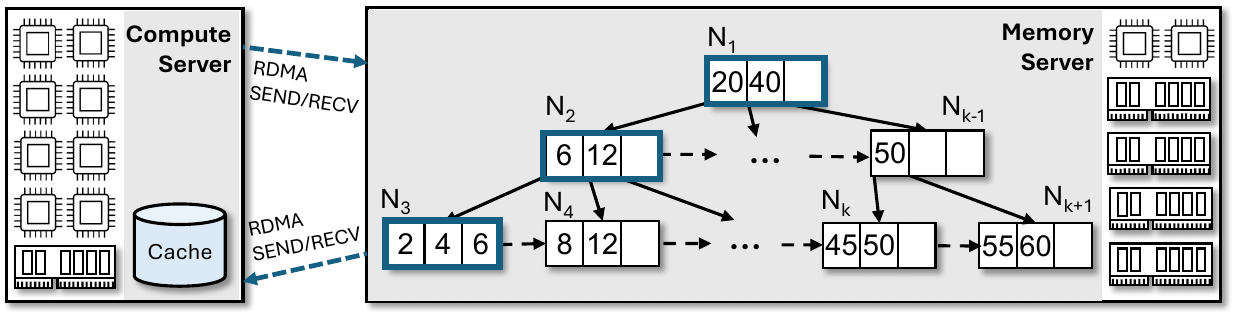}\vspace{-.05in}
        \subcaption{Two-sided RDMA indexing.}\vspace{-.15in}
        \label{fig:indexing_scenario_twosided}
    \end{minipage}
    \caption{Indexing scenarios of read operations ($key=4$) in disaggregated memory databases.}\vspace{-.1in}
    \label{fig:indexing_scenario}
\end{figure*}

\subsection{RDMA}
\label{sec:rdma}

RDMA network provides higher bandwidth and lower latency than conventional TPC/IP networks. 
Moreover, it enables direct access to remote memory without remote CPU involvement.
% Moreover, RDMA provides the unique characteristic of direct access to remote memory without remote CPU involvement.
These features have opened up new opportunities for optimizing distributed database systems. 
Research on RDMA can be divided into two directions, enhancing traditional RPC-based systems with two-sided RDMA and redesigning the systems to leverage one-sided RDMA.
% Research on RDMA can be divided into two directions, optimizing traditional RPC-based systems with two-sided RDMA and redesigning the systems to leverage one-sided RDMA.
% Research directions with RDMA can be divided into two categories, optimizing traditional RPC-based systems with two-sided RDMA, and redesigning the systems to leverage one-sided RDMA.

\textit{Two-sided RDMA} provides a lightweight message-passing abstraction with a SEND/RECV interface, commonly used to implement RPCs~\cite{scaleRPC, herd, fasst, eRPC}.
A message is exchanged when a SEND matches a RECV.
On the receiver side, the host must poll for incoming requests and manager the receive buffer.
% On the receiver side, the host must poll the message request to successfully receive it and manage the buffer contents.
Therefore, two-sided RDMA involves the CPUs on both the sending and receiving ends, not providing the CPU-bypassing property as in one-sided RDMA.

\textit{One-sided RDMA} registers the memory region of a remote host in its local NIC and provides direct access to remote memory through READ/WRITE/ATOMIC interfaces.
RDMA ATOMIC supports Fetch-And-Add (FAA) and Compare-And-Swap (CAS) at 8-byte granularity, which are commonly used to implement synchronization protocols.
Since one-sided RDMA does not involve remote CPUs, it has been considered as a well-fitting communication method in disaggregated systems~\cite{rolex, sherman, namdb, namtree}.

% \subsection{RDMA-based Transaction Processing}
\subsection{RDMA-based Transaction Processing in Disaggregated Architecture}
\label{sec:rdma_txn}

Transaction processing in existing disaggregated systems mainly works with one-sided RDMA.
This preference is largely because one-sided RDMA can leverage strong computational power in compute servers to directly access remote memory in memory servers, 
while two-sided RDMA requires involvement of limited CPU resources in memory servers~\cite{sherman, rolex, dsmdb, namdb}.
% Therefore, with one-sided RDMA, indexes and tables are stored and maintained in memory servers, and their memory regions are registered in each compute server, allowing global shared memory access.
We break down the transaction processing into indexing and concurrency control and describe how they work with RDMA in the two memory disaggregation models, non-partitioned memory and co-partitioned memory.\vspace{-.1in}

\subsubsection{Indexing}
\label{sec:rdma_index}

While a lot of prior work has studied RDMA-based indexing, most designs are hash-based~\cite{pilaf, farm, rcc, race, active_memory}.
This is because one-sided RDMA indexing incurs multiple network transfers, and hashing bounds the number of remote memory accesses by limiting bucket probes~\cite{cuckoo}.
% This is because one-sided RDMA requires data transfer over the network multiple times, and hashing bounds the number of remote memory accesses by limiting the number of bucket accesses~\cite{cuckoo}.
However, hash indexes have a critical limitation that they do not support range scans and thus cannot be used in many practical workloads.
B+-tree-based RDMA indexes, on the other hand, support various workload requirements but suffer from network amplification, as index traversal requires multiple RDMA READs to locate each node from the root to the leaf.
% The index traversal requires multiple RDMA READs, to locate each node from the root to the leaf.

Figure~\ref{fig:indexing_scenario_onesided} illustrates an indexing scenario in one-sided RDMA B+-tree.
The compute server retrieves the value for key 4 by traversing the tree in the memory server.
Given an index height of 3, the compute server performs at least 3 RDMA READs, i.e., $N_1$, $N_2,$ and $N_3$, to reach the target leaf.
For each access, it reads the entire node data, typically 1~KB in size~\cite{namtree, sherman, dex}, to locate the child.
Although omitted from the figure for clarity, each node access requires at least three dedicated RDMA READs to ensure correctness in optimistic synchronization~\cite{rdma_guideline}, i.e., reading the node version, reading the node data, and validating the version.
% Moreover, each node access requires at least three dedicated RDMA READs to guarantee the correctness in optimistic synchronization~\cite{rdma_guideline}, i.e., reading a version of the node, reading the node data, and validating the version.
Overall, these RDMA READs for traversal and synchronization cause significant network amplification, which gets worse as the index grows.

% To minimize the cost of tree traversal, $B^{link}$-tree~\cite{blink1981} has been conventionally used in disaggregated systems.
% It allows a compute server to jump to its right sibling node in the middle of the traversal when an update on the node has been detected, instead of retrying.
% It employs a version-based optimistic latch coupling~\cite{olfit2001} for synchronization to avoid any updates in the global shared memory during the traversal.
% To fully leverage one-sided RDMA, each index node is distributed in the memory region of a memory server.
% Upon an indexing request, a compute server identifies the memory address of a data tuple by traversing the index.

% Index traversal requires multiple RDMA READs, to locate each node from the root to the leaf.
% This results in significant network amplification, which gets worse as the index grows.
% Moreover, a recent study has pointed out the limitation of optimistic synchronization with one-sided RDMA that RDMA operations may corrupt data if they are not correctly serialized~\cite{rdma_guideline}.
% That is, overlapping multiple RDMA operations may result in incorrect results due to operation reordering in PCIe, and RDMA READ does not guarantee reading the memory address in ascending order as done in RDMA WRITE.
% Therefore, each node access during index traversal requires at least three dedicated RDMA READs to read the version of a node, read the node data, and validate the version, which exacerbates the network amplification problem.

To mitigate the network amplification problem, prior work has employed caching in compute servers to reduce remote accesses during index traversal~\cite{sherman, deft, dex}.
In non-partitioned memory, multiple compute servers may cache the same data since they have shared access to memory servers.
To keep cache coherence simple, recent studies cache shortcut hints to data tuples, e.g., last-level nodes in a B+-tree~\cite{sherman, deft}.
Then, the system uses a cached hint to locate remote data and validates staleness by checking its metadata.
In co-partitioned memory, each compute server owns a logical data partition in memory servers and caches only the index nodes in its partition, avoiding coherence management across compute servers.
% In co-partitioned memory, compute servers are partitioned to avoid cache coherence across multiple compute servers.
% That is, each compute server logically owns a specific data partition in memory servers and caches tree nodes that belong to the partition~\cite{dex}.

% Therefore, cache coherence only happens within a compute server.
% \yxy{I feel we can remove the last sentence since it causes a minor confusion} \hcha{I commented it, but can i ask what the confusion is?} \yxy{intuitively, I would think of it as having no cache coherence problem. So when you say CC within a compute server, I wonder what it means.. Then I realize whatever it means does not really matter}

% However, this increases the system complexity due to cache coherence management since data in memory servers are shared across compute servers.
% One approach is to cache shortcut hints to data tuples, e.g., last-level nodes in a B+-tree, and postpone the cache coherence to later data accesses~\cite{sherman, deft}.
% Upon a data access, the system uses a cached hint to locate the remote data and validates its staleness by checking its metadata.
% Another approach is to partition compute servers to avoid unnecessary cache invalidation~\cite{dex}.
% Each compute server logically owns a specific partition of data and caches tree nodes that belong to the partition such that cache coherence happens within the partition, instead of across partitions.

While these techniques have shown effectiveness in one-sided RDMA indexing, they have not been explored in the context of two-sided RDMA indexing.
We make a key observation that the caching approaches can also be applied to two-sided RDMA systems.
In non-partitioned memory, shortcut hints can be cached in compute servers and sent to memory servers via network messages to reduce indexing overhead.
In co-partitioned memory, each compute server can cache data tuples that belong to its own partition, completely eliminating network round-trips for cached data retrieval.
% Moreover, compute-side logical partitioning allows each compute server to maintain its local cache without coherence management across compute servers.

\subsubsection{Concurrency Control}
\label{sec:rdma_cc}
% \yxy{this section is kinda long. making it a bit shorter would be nice.}\hcha{made it shorter. please check if it looks  okay.}
% Different concurrency control protocols can be applied to ensure the correctness of concurrent transactions in disaggregated memory databases.
In co-partitioned memory, each compute server accesses its own partition in memory servers, allowing concurrency control to be handled locally in compute servers.
% Therefore, concurrency control can be performed locally in compute servers without remote coordination.
However, this requires an additional distributed transaction layer for cross-partition transactions on top of compute servers, e.g., proxy servers or client servers.
% However, this requires a distributed transaction protocol to handle cross-partition transactions in an additional layer on top of compute servers, e.g., proxy servers or client servers.
In non-partitioned memory, compute servers have shared access to memory servers, requiring a scalable concurrency control protocol to manage concurrent accesses.
% Since compute servers have shared access to memory servers over one-sided RDMA, their concurrent accesses to the memory pool need to be protected through a scalable concurrency control protocol.

Optimistic concurrency control (OCC)~\cite{occ1981} and two-phase locking (2PL)~\cite{2pl1979, 2pl1981} are the two most widely-used protocols in RDMA-based systems~\cite{polardb, farm, drtmh, drtm, active_memory, chiller}.
OCC defers conflict detection to the end of execution by validating versions of accessed records, while 2PL synchronizes via explicit lock acquisition in a shared or exclusive mode.
% In OCC, transactions postpone conflict detection to the end of execution by validating versions of accessed records.
% In contrast, 2PL synchronizes transactions through explicit shared or exclusive lock acquisition.
Although both are simple and efficient, OCC lacks native support for priority-aware conflict handling, causing starvation prevention problems.
Recent work has sought to reduce high tail latency in OCC by incorporating pessimistic mechanisms from 2PL~\cite{polaris, f1}, highlighting the importance of lock-based control under contention.
Therefore, we focus on 2PL in this work.
% While both schemes are simple and efficient, we focus on 2PL protocols in this work since OCC lacks priority consideration.

To implement 2PL with one-sided RDMA, an 8-byte value is sliced into two smaller values to indicate the number of lock owners for exclusive and shared access, respectively.
Acquiring a lock requires at least two RDMA round-trips.
% The lock value is first read with RDMA READ to check for conflicts, and the value is atomically updated with RDMA CAS if there is no conflict.
First, the lock value is read with RDMA READ to check for conflicts.
% It will abort upon a conflict.
% Otherwise, the value is atomically updated with RDMA CAS.
If no conflict exists, the value is then atomically updated with RDMA CAS.
% It will abort upon a conflict.
% Otherwise, the value is atomically updated with RDMA CAS.

A critical limitation of 2PL with one-sided RDMA is that it cannot guarantee starvation prevention.
In conventional 2PL, \waitdie and \woundwait prevent starvation by scheduling transactions based on priorities and allowing higher-priority transactions to wait for or preempt lower-priority ones.
% In conventional 2PL protocols, \waitdie and \woundwait provide starvation prevention by scheduling transactions based on priorities and allowing a transaction to preempt another transaction that has lower priority.
However, the limited APIs of one-sided RDMA make it extremely difficult to implement such functionalities.
% Priority-based scheduling requires transferring the ownership of a lock to the next waiting transaction with the highest priority.
Priority-based scheduling requires transferring lock ownership to the next waiting transaction in priority order, which in turn demands a notification mechanism among compute servers.
% To provide priority-based scheduling, the ownership of a lock needs to be handed over to the next waiting transaction with the highest priority.
% This requires a notification strategy to the transaction which needs additional communication mechanism among compute servers.
Similarly, preemption requires notifying transactions of their preempted status.
% Such notification is also required in preemption, to notify the transactions of their preempted status. 
% To provide preemption, the system needs to maintain the status of transactions to prevent committing transactions from being preempted.
% Once preempted, it requires another strategy to notify preempted transactions of their status.
Neither of the protocols can be efficiently implemented with one-sided RDMA since all these operations need to be executed in an atomic way.

% \hcha{if the section needs further shrinking, this paragraph can be removed.}
One way to reduce latency is to place an additional 8-byte value alongside a lock to track the priority of a current lock owner~\cite{rcc}.
After lock acquisition, the owner updates its priority value via RDMA CAS, allowing other transactions to decide whether to wait or abort upon a conflict.
% After lock acquisition, the owner updates the priority value of the lock with its transaction priority via RDMA CAS, letting other transactions determine to wait or abort upon a conflict.
However, this still does not guarantee starvation prevention.
% When a lock is released, its ownership is not guaranteed to be handed over to the next waiter with the highest priority.
Since a lock should be acquired via RDMA CAS, whichever transaction succeeds in swapping the lock value becomes the next owner, regardless of priority.

While these limitations of one-sided RDMA remain unsolved, two-sided RDMA can easily support those functionalities with the help of remote CPUs.
As transaction requests are transferred to memory server CPUs over network messages, the system can leverage optimized legacy data structures and algorithms to implement starvation prevention, priority-based scheduling, and preemption.
\section{\work Overview}
\label{sec:overview}

% \yxy{don't pitch this section as a overview. This section is talking about Lotus design itself. Sec 4 and 5 are about Lotus' key optimizations, caching and batching. Not sure whether you want to add another section later on CC, if not, then CC is described mainly in section 3, so it's certainly not just an overview. With this in mind, in Section 3.1, let's try to talk about everything about indexing except the optimizations that we will discuss in later sections. (I think the current contents are doing this already, just adjust the tone a bit)}
% \hcha{Changed the name of this section to Overview, and revised the contents. Please check.
% I also changed the organization to cover CC first in Section 4, and then talk about optimization techniques in Sections 5 and 6, because I think it makes a better flow (discussing CC before batching).}

We introduce \work, a scalable transaction processing system for disaggregated DBMSs that supports both non-partitioned and co-partitioned memory.
It leverages the rich semantics with two-sided RDMA for indexing and concurrency control and further optimizes the processing pipeline through caching and batching.
% We introduce \work, a scalable transaction processing system for disaggregated DBMSs that employs two-sided RDMA for remote access.
% \yxy{emphasize it's 2-sided}
% Unlike in many two-sided RDMA systems~\cite{fasst, herd, scaleRPC}, \work uses reliable connection (RC) queue pairs (QPs) for network communication between compute servers and memory servers.
% \work leverages rich semantics for indexing and concurrency control, and further optimizes the processing pipeline through caching and request batching.
In this section, we describe the design of indexing and concurrency control in \work and provide an overview of our optimization techniques.
\noindent\textbf{Indexing.}
Unlike in many one-sided RDMA indexes~\cite{sherman, deft, dex, rolex, smart, namtree}, which maintain a global index in memory servers, 
\work maintains a local index in each memory server.
% a DBMS is partitioned across memory servers, and each of them maintains a local index for its partition.
% \yxy{are you trying to explain how Lotus works or how DBMSs work in general? some clarification can help here.} \hcha{Made it clear.}
An indexing operation is executed by sending a request from a compute server to the corresponding memory server over two-sided RDMA.

The limitation of one-sided RDMA indexing, i.e., network amplification, can be easily addressed with two-sided RDMA, as it offloads index traversal to memory server CPUs.
Processing an index request needs at most one network round-trip with minimal traffic.

Figure~\ref{fig:indexing_scenario_twosided} shows an indexing scenario using two-sided RDMA.
The memory server maintains a B+-tree index, and the compute server is trying to retrieve the value for a key.
The compute server first sends the index request to the memory server via RDMA SEND, which is only a few bytes in size.
The memory server then traverses the index by directly accessing its local memory and returns only the result to the compute server.
The cache in the compute server is used to reduce the overhead of network and data processing.
\noindent\textbf{Concurrency control.}
Similarly to indexing, the limitation of one-sided RDMA concurrency control, i.e., functional deficiency, can be effectively addressed with two-sided RDMA by offloading lock management to memory server CPUs.

\work maintains lock tables in each memory server and implements critical system functions in concurrency control protocols, including starvation prevention, priority-based scheduling, and preemption.
We discuss the technical details in Section~\ref{sec:concurrency_control}.
% \yxy{Given the current 3.1 and 3.2 are quite simple, maybe they don't need to be subsection, but something like the following \\
% \noindent\textbf{Indexing.} xxx\\
% \noindent\textbf{Concurrency Control.} xxx
% }
% \hcha{I also modified Optimization part to follow the same layout. I will undo it if optimization technique is worth having a subsection.}

% \subsection{Optimization Techniques}
% \label{sec:optimization}

% \yxy{I think we can downplay the discussion a bit on the limited compute resources for 2-sided. We have been emphasizing this quite a bit throughout the paper. With this in mind, no need to call this section "Challenges in Lotus"; we are providing an overview of Lotus' optimizations that we will discuss in later sections.}
% \hcha{Revised the discussion.}
\noindent\textbf{Optimization techniques.}
While the adoption of two-sided RDMA provides rich system functionality for indexing and concurrency control, the involvement of memory server CPUs may introduce performance overhead.
To address this challenge, we present two key optimization techniques that minimize the overhead of network and data processing.
These optimizations can significantly reduce the number of remote data accesses and network round-trips.\vspace{-.05in}
\begin{itemize}[leftmargin=*]
    \item \textbf{Caching.} \work adopts the state-of-the-art caching techniques discussed in Section~\ref{sec:rdma_index} for non-partitioned memory and co-partitioned memory to accelerate indexing. We enhance these techniques by converting them to be compatible with two-sided RDMA. We describe the details in Section~\ref{sec:cache}.
    % \item \textbf{Caching.} \work adopts the two state-of-the-art caching techniques discussed in Section~\ref{sec:rdma_index}, shortcut hint caching and compute-side partitioned caching, to accelerate indexing. We enhance these techniques by converting them to be compatible with two-sided RDMA. We describe the details in Section~\ref{sec:cache}.
    \item \textbf{Batching.} \work applies two batching techniques, intra-transaction batching and inter-transaction batching, to minimize the number of network round-trips in transaction execution. We discuss these techniques in Section~\ref{sec:batch}.

\end{itemize}

\section{Concurrency Control}
\label{sec:concurrency_control}

\begin{figure*}[!t]
    \centering
    \begin{minipage}{0.3\textwidth}
        \centering
        \includegraphics[width=\textwidth]{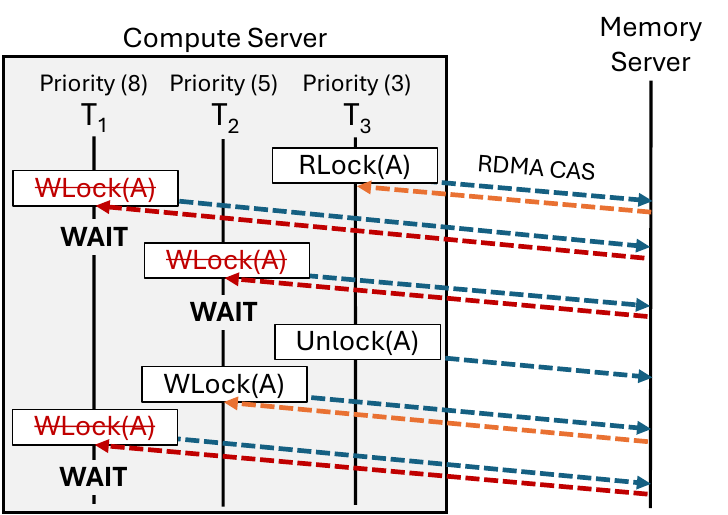}\vspace{-.05in}
        \subcaption{One-sided RDMA \waitdie.}\vspace{-.1in}
        \label{fig:cc_onesided_waitdie}
    \end{minipage}
    \hfill
    \centering
    \begin{minipage}{0.3\textwidth}
        \centering
        \includegraphics[width=\textwidth]{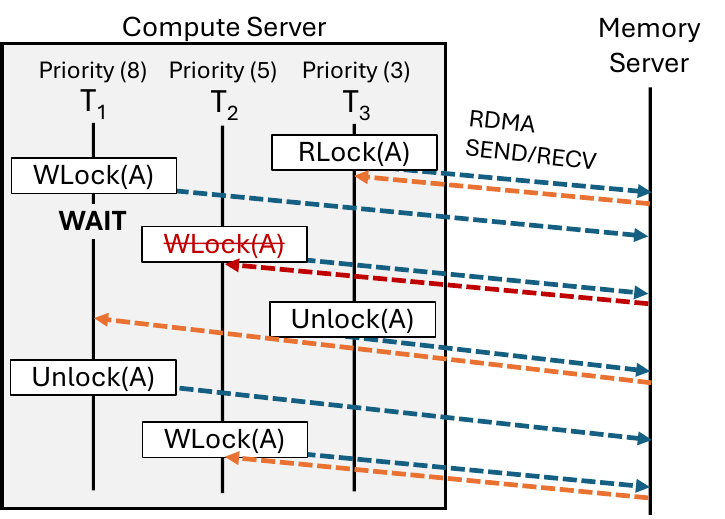}\vspace{-.05in}
        \subcaption{Two-sided RDMA \waitdie.}\vspace{-.1in}
        \label{fig:cc_twosided_waitdie}
    \end{minipage}
    \hfill
    \centering
    \begin{minipage}{0.3\textwidth}
        \centering
        \includegraphics[width=\textwidth]{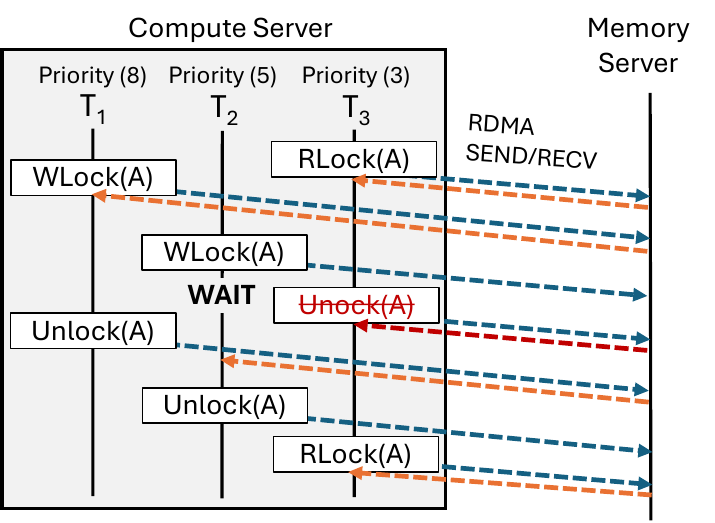}\vspace{-.05in}
        \subcaption{Two-sided RDMA \woundwait.}\vspace{-.1in}
        \label{fig:cc_twosided_woundwait}
    \end{minipage}
    \caption{Execution scenario of two write transactions ($T_1$ and $T_2$ with priorities 8 and 5) and one read transaction ($T_3$ with priority 3) --- higher values represent higher priorities. Blue arrows denote lock requests, orange arrows represent successful lock acquisitions, and red arrows indicate lock acquisition failures.}\vspace{-.1in}
    \label{fig:cc_scenario}
\end{figure*}

% Concurrency control in co-partitioned memory is straight-forward as it can be performed locally in compute servers without RDMA accesses.
% Therefore, we focus on the concurrency control in non-partitioned memory in this section.
% \hcha{Reviewer 1 was confused if \work supports non-partitioned memory, co-partitioned memory, or both. I added a sentence in Section~\ref{sec:overview}, but do you think we should also clarify that we focus on concurrency control for non-partitioned memory as it }
Concurrency control with two-sided RDMA enables rich functionalities that are not supported with one-sided RDMA, as discussed in Section~\ref{sec:rdma_cc}.
By offloading lock management to memory server CPUs, the system can use optimized data structures and algorithms to implement starvation prevention, priority-based scheduling, and preemption.
% The system offloads lock management to memory server CPUs and can utilize the CPUs to implement various data structures and algorithms to support starvation prevention, priority-based scheduling, and preemption.
Compute servers send lock requests to memory servers over two-sided RDMA, and memory servers process these requests by directly accessing their local lock tables.
% The limitation of one-sided RDMA concurrency control, i.e., functional deficiency, can be addressed with two-sided RDMA by offloading the lock management to memory server CPUs.
% \hcha{We first review one-sided RDMA concurrency control and describe how \work provides the missing functionalities with an example. (please check)}
We first describe how \work implements these functionalities and then walk through an example to provide a comprehensive understanding of the details.
% We first walk through an example of one-sided RDMA concurrency control to review its challenge, i.e., functional deficiency, and then, describe how it is resolved in two-sided RDMA.

% \subsection{Implementation Details}
% \label{sec:implementation_details}

\subsection{Priority-based Scheduling}
% \noindent\textbf{Priority-based scheduling.}
\label{sec:priority_based_scheduling}
% \yxy{seems there are quite a lot of details in CC. Maybe it makes sense to move CC to an entirely new section (like Section 6). If you do that, Section 3 actually acts like an overview---but given Sections 4 and 5 are optimizations, we still want to have the tone that Section 3 is describing the key design, instead of acting as a preview of later sections.} 
% \hcha{I moved CC to a new section, but then, Section 3 does not have enough contents to be considered as a design section.}
% To achieve low latency and fairness,
To achieve low latency and ensure fairness among concurrent transactions,
\work implements priority-based scheduling in memory servers.
Each memory server maintains a wait queue per lock to manage conflicting requests in priority order.
% \yxy{wait-for graph is for deadlock detection in a DAG, which is not what we have here. We are maintaining a waiting queue for each lock.}\hcha{fixed it.}
Upon a conflict, the lock manager compares the priority of the incoming request with that of the lock owners.
Based on the concurrency control protocol, the manager may add the request to the wait queue, preempt lower-priority owners, or reject the request.
% \yxy{the manager may also reject the incoming request in wait-die}\hcha{added rejection case}
When a lock is released, its ownership is granted to the waiting transaction according to the priority order, ensuring starvation prevention.
% \yxy{not necessarily true. In wait-die, high-priority txn wait for Low-p txn, if a wait-queue has multiple txns, the manager takes the one with the lowest priority---otherwise you violate high-p waiting for low-p invariant. I suggest changing it to "ownership is granted to the waiting transactions according to the priority order" } \hcha{You are right. I changed it}

\subsection{Preemption}
% \noindent\textbf{Preemption.}
\label{sec:preemption}
Preemption is another critical functionality in concurrency control protocols that forces lower-priority transactions to abort.
% Preemption is another critical functionality in concurrency control protocols by forcing to kill active transactions that have lower priorities.
We implement preemption by maintaining an additional structure that tracks transaction status.
Upon a lock conflict, the lock manager checks the lock table and attempts to preempt lower-priority owners by removing them from the table and atomically updating their status.
This process is protected by a mutex to ensure atomicity.
If all conflicting transactions have been preempted, the lock is granted to the incoming request.
Some transactions may not be preempted if they have higher priorities or have already entered a commit phase.
Then, the request is added to the wait queue.

\subsection{Lazy Notification}
% \noindent\textbf{Notification}
\label{sec:notification}
Transactions may have to wait for lock acquisition or be preempted due to conflicts, as discussed above.
These functions may increase the complexity of communication between compute servers and memory servers.
In two-sided RDMA, message delivery requires a matching pair of SEND and RECV.
Since a transaction may need extra waiting time to acquire locks and be preempted at arbitrary points, it may lead to a mismatch of the two.
For example, after receiving a lock grant, a transaction may post another SEND to a memory server to request additional locks.
However, if the previously granted lock has been preempted, the memory server must notify the preempted compute server transaction to abort by posting a SEND, leading to a SEND-RECV mismatch.

To handle the mismatch problem in a simple and efficient way, \work notifies lock status to compute servers in a lazy manner.
% To keep it simple and efficient, \work handles such cases by notifying the status to compute servers in a lazy manner.
% In such cases, \work does not immediately respond to a compute server about the status.
% Instead, it notifies the compute server in a lazy manner.
For a waiting transaction, a memory server does not immediately notify the compute server about the waiting status. 
% \yxy{use "the memory server" instead of \work?} 
Instead, the memory server keeps the lock request in the wait queue, periodically checking for lock availability, and lets the compute server wait.
Once the lock becomes available, it processes the request and then notifies the compute server.
% \yxy{it still says "sends result back to compute server"}
For a preempted transaction, the memory server simply marks its status as preempted without immediate notification.
When the next request from the preempted transaction arrives, the memory server informs that the transaction has been preempted, forcing to abort.
This lazy notification further eliminates unnecessary network communication between servers.
% between compute servers and memory servers.
% \yxy{I think a challenge that leads to this design is that 2-sided communication must be initiated by the compute server, not the memory server. Not sure whether we should mention this somehow.} \hcha{Added a paragraph about it in 4.3.}

\subsection{Example}
\label{sec:example}
% \subsubsection{Challenges in One-sided RDMA.}
% \yxy{does it make sense to fold the contents here into Sec 2.3.2? Right now we discuss the limitations 1-sided CC in two different places.}
% \hcha{If we do that, Sec 2.3.2 might get too long given it is already kinda long. I changed the tone of this subsection to first review one-sided CC for a clear comparison with our CC, instead of emphasizing the limitation.}

Now, we illustrate how \work guarantees starvation freedom through priority-based scheduling, preemption, and lazy notification.
% Now, we walk through an example to provide a comprehensive understanding on how priority-based scheduling, preemption, and lazy notification in \work guarantee starvation freedom.
For a clear comparison, we first review one-sided RDMA concurrency control.
Figure~\ref{fig:cc_scenario} shows execution scenarios of two write transactions, $T_1$ and $T_2$, and one read transaction, $T_3$, all accessing the same tuple $A$.
$T_1$ has the highest priority, followed by $T_2$ and $T_3$.
We assume the following execution order: $T_3 \rightarrow T_1 \rightarrow T_2$.
For clarity, we omit metadata and data accesses and focus on lock operations.

Figure~\ref{fig:cc_onesided_waitdie} shows the execution scenario in one-sided RDMA \waitdie protocol, described in Section~\ref{sec:rdma_cc}.
$T_3$ first acquires a read lock via RDMA CAS.
$T_1$ tries to acquire a write lock but fails due to a conflict.
Then, $T_1$ compares its priority with that of the current owner, $T_3$.
Since $T_1$ has higher priority, it waits until $T_3$ completes.
Meanwhile, $T_2$ also fails to acquire the lock and waits.
The waiting in one-sided RDMA involves periodic RDMA READs to poll the lock availability.
Once $T_3$ releases the lock, both $T_1$ and $T_2$ try to acquire it.
However, $T_2$ succeeds before $T_1$ because lock acquisition via RDMA CAS does not enforce ordering.
Therefore, $T_1$ has to keep waiting until $T_2$ completes, which may suffer indefinitely.
% This shows the fundamental limitation of one-sided RDMA concurrency control that it does not guarantee starvation prevention.

% \subsubsection{Two-sided RDMA Solutions.}
% The limitation of one-sided RDMA concurrency control can be addressed with two-sided RDMA by offloading the lock management to memory server CPUs.

Figure~\ref{fig:cc_twosided_waitdie} describes the same scenario in two-sided RDMA \waitdie.
$T_3$, $T_1$, and $T_2$ send their lock requests to the memory server over RDMA SEND.
Since $T_3$'s request has arrived first, it acquires the lock.
Then, the memory server schedules the high priority transaction, $T_1$, in the wait queue and informs $T_2$ to abort since $T_2$ has lower priority than the waiting transaction, $T_1$.
Note that $T_1$ is notified when the lock is granted, as discussed in Section~\ref{sec:notification}.
% We discuss how we manage requests in priority order in Section~\ref{sec:implementation_details}.
When $T_3$ completes, the memory server hands over the lock to $T_1$.
After $T_1$ completes, $T_2$ retries and acquires the lock.
% \yxy{WAIT\_DIE does not allow a low-priority txn to wait for a high-priority txn. So if T1 is waiting for T3, T2 cannot join the wait list, since T2 cannot wait for T1. Instead, T2 will abort and retry later.} \hcha{Right. I fixed the figures and edited this paragraph.}

Figure~\ref{fig:cc_twosided_woundwait} illustrates the two-sided RDMA \woundwait case.
$T_3$ first acquires the lock.
However, after $T_1$ sends its request, the memory server preempts $T_3$ and grants the lock to $T_1$ since $T_1$ has higher priority.
% However, after the memory server receives the request from $T_1$, it preempts $T_3$ and grants the lock to $T_1$ since $T_1$ has higher priority.
Note that $T_3$ is not notified yet about its preempted status, as discussed in Section~\ref{sec:notification}.
% We discuss how we implement the preemption in Section~\ref{sec:implementation_details}.
After receiving the request from $T_2$, the memory server adds it to the wait queue.
Then, when $T_3$ tries to commit, the memory server notifies $T_3$ to abort as it keeps track of $T_3$'s preempted status. % of preempted transactions, i.e., $T_3$ in this example.
After $T_1$ completes, the lock is granted to the next waiter, $T_2$.
When $T_3$ retries, its request is added to the wait queue since the current lock owner, $T_2$, has higher priority.

\section{Caching}
\label{sec:cache}
Caching can significantly reduce both remote data processing and network processing overhead.
Based on the memory disaggregation models that handle cache coherence in different ways, as discussed in Section~\ref{sec:motivation}, we adopt two caching techniques for both (1) non-partitioned memory and (2) co-partitioned memory.
Our caching techniques operate at a finer granularity than conventional node-level caching, i.e., record-level caching.
% Based on the memory disaggregation models that handle cache coherence in different ways, as discussed in Section~\ref{sec:motivation}, we adopt two caching techniques that operate at a finer granularity, i.e., record-level caching, than conventional node-level caching, for both (1) non-partitioned memory and (2) co-partitioned memory.
% \yxy{break up this long sentence} \hcha{edited.}
% Caching can significantly reduce remote data processing and network processing overhead, but cache coherence introduces new challenges, as discussed in Section~\ref{sec:motivation}.
% To address these challenges, we adopt two caching techniques that handle coherence under different memory models and at a finer granularity than conventional node-level caching, i.e., record-level caching, for (1) non-partitioned memory and (2) co-partitioned memory.
% We apply two caching techniques that handle the coherence in two different ways: 
% (1) shortcut hint caching~\cite{sherman, deft} in non-partitioned memory and 
% (2) compute-side partitioned caching~\cite{dex} in copartitioned memory.
In the following, we describe how each technique is applied to our indexing design.
% In the following, we describe how we adopt these two techniques to our indexing.
% \hcha{I changed the name of subsections (Shortcut Hint Caching --> Caching for Non-Partitioned Memory, and Compute-Side Partitioned Caching --> Caching for Copartitioned Memory). I think the previous names tend to make an impression that we are doing just the same as previous work. }

\subsection{Caching at Record Granularity}

Unlike prior one-sided RDMA indexing schemes that cache index nodes, \work caches individual records.
Caching index nodes naturally complements one-sided RDMA, where compute servers directly traverse index structures in memory servers.
In two-sided RDMA, compute servers send requests and receive responses, while memory servers handle index traversal internally on their behalf.
This execution model not only preserves abstraction boundaries but also enables effective record-level caching along the request-handling path, which is difficult to achieve with one-sided RDMA.

Record-level caching also improves cache utilization by providing finer-grained control over data access.
Recent studies have shown that caching index nodes may lead to inefficient cache utilization since a single node contains many records and only a few of them may be frequently accessed~\cite{twotree, anticaching, siberia}.
% Motivated by these observations, \work implements a cache for two-sided RDMA that operates at the level of individual records.

Cache replacement in \work is also performed at the record level.
\work tracks the access frequency of each cache record and evicts the least frequently accessed one when the cache becomes full.
% To further improve cache efficiency, \work employs probabilistic admission.
% When cache size is limited, frequent admissions may lead to excessive replacements, causing newly admitted records to be evicted before they accumulate sufficient access frequency.
% Probabilistic admission mitigates this issue by selectively admitting new records, allowing them to build access counts, while preserving frequently accessed records in the cache.

\subsection{Caching for Non-Partitioned Memory}
% \subsection{Shortcut Hint Caching}
\label{sec:shortcut_hint_caching}

As discussed in Section~\ref{sec:rdma_index}, multiple compute servers may cache the same data in non-partitioned memory.
To keep cache coherence management simple, \work follows the approach from prior work~\cite{sherman, deft} that caches pointers to data tuples.

In non-partitioned memory caching, read and write requests follow the same handling path.
A compute server sends a request to a memory server, and the memory server traverses its index with the request key and retrieves the pointer to the corresponding data.
Along with the data, the memory server returns this pointer, which is then cached in the compute server.
On subsequent accesses to the matching key, the compute server provides the pointer to the memory server as a shortcut hint.
This allows the memory server to directly access the data, bypassing index traversal.

% Shortcut hint caching~\cite{quit, postgresql}
% \yxy{is there a good reference for it?}\hcha{not perfectly matching references, but added two that I think related. PostgreSQL uses a pointer to the rightmost leaf node (tail leaf) to accelerate insertion for sorted data to avoid B+-tree traversal, and QuIT extends it by maintaining multiple pointers (e.g., tail leaf, last-insertion leaf, and predicted leaf from a model)} 
% reduces data processing overhead in memory servers for non-partitioned memory systems where a compute server can access any partition of the DBMS in memory servers.
% Therefore, multiple compute servers may cache the same data, which requires cache coherence management.
% In this caching, compute servers cache pointers to data tuples in its local memory, instead of caching the actual tuple data, to keep cache coherence management simple.
% When a compute server sends a request to a memory server, the memory server traverses an index with the given key and retrieves the pointer to the data tuple.
% Instead of just returning the tuple data, the memory server also provides the pointer.
% Then, the compute server caches it in local memory.
% Upon an access to the matching key in the cache, it provides the pointer to the memory server as a shortcut hint.
% The memory server, then, directly accesses the tuple with the provided hint, avoiding index traversal.

Some cache entries may become stale if the original data in a memory server has been updated by requests from other compute servers.
Validation of such stale cache entries is offloaded to memory servers.
When a shortcut hint is provided, the memory server checks tuple metadata to determine its validity.
If stale, the memory server retrieves the up-to-date data via index traversal and returns it to the compute server.
The compute server then invalidates the stale cache entry and updates it with the new one.

Cache replacement may occur upon admission if the cache becomes full.
Since cached data serves only as a hint to accelerate remote data access, eviction does not require coherence management.
The compute server simply removes the entry from the cache.
% We discuss our replacement policy in Section~\ref{sec:cache_replacement}.

% \yxy{should we also discuss write for non-partitioned memory? (we discuss writes for co-partitioned memory below.)} \hcha{I did not include discussion about writes explicitly, because cache for non-partitioned memory does not differentiate reads and writes. It's just used as a shortcut to bypass index traversal in memory servers. When writes update the data in memory servers, cache may become stale, which can be captured in the validation process discussed in the third paragraph. Do you think we should talk about this?}\yxy{makes sense. might help if add a short explanation, like in non-partitioned arch, caching for reads and writes are handled in the same way.}\hcha{Added the sentence.}
% \yxy{From reading this subsection, it is not crystal clear what is prior work vs. new idea in Lotus. It feels everything is prior work. If possible, better highlight what is new in Lotus.}\hcha{I added 5.1 to highlight the difference.}

% This caching mechanism can efficiently reduce the overhead of indexing while keeping the cache coherence management simple.
% The cache is only updated when a cache miss or a stale cache access occurs, minimizing the number of coherence operations.
% However, it must involve remote memory accesses whether it is a cache hit or miss, since concurrency control is required in memory servers to ensure correct synchronization among compute servers.

\subsection{Caching for Co-Partitioned Memory}
% \subsection{Compute-Side Partitioned Caching}
\label{sec:compute_side_partitioned_caching}

% Compute-side partitioned caching, on the other hand, targets co-partitioned memory systems, which handles cache coherence in a different way.
% Unlike in non-partitioned memory systems, co-partitioned memory systems additionally partition the DBMS in the compute layer.
% This eliminates the need for cache coherence across compute servers since each of them only accesses its own partition.
% Therefore, compute-side partitioned caching caches actual tuple data instead of pointers to tuple data.

% Prior work on one-sided RDMA indexing~\cite{dex} caches the index nodes that are not present in the cache during tree traversal to reduce remote memory accesses.
% While this approach provides a natural indexing case in one-sided RDMA as compute servers directly modify the index in memory servers, it does not provide huge benefits in two-sided RDMA because the indexing path is not usually exposed to compute servers.
% Therefore, \work caches individual records instead of index nodes in co-partitioned memory systems.
% A recent work has demonstrated that caching at the granularity of index nodes may lead to inefficient cache utilization,
% since multiple records are stored in a single index node, and only a few of them may be frequently accessed~\cite{twotree}.
% We follow this observation and cache individual data tuples in compute servers to improve cache efficiency.

Unlike caching for non-partitioned memory, caching for co-partitioned memory stores actual tuple data, since cache coherence does not occur across compute servers.
% Different from shortcut hint caching, compute-side partitioned caching in \work stores actual tuple data in cache.
% This design simplifies the system by eliminating the need for cache coherence across compute servers.
When a compute server processes a request, it first checks its local cache.
On a cache miss, the compute server requests the data from the memory server and caches it locally upon return.
On a cache hit, the compute server accesses the data directly from the cache without involving the memory server.
% Upon a cache miss, the compute server sends the request to the memory server to access the remote data.
% Once the data is returned, it is cached in the compute server for future accesses.
% Upon a cache miss, it latches the request key to prevent other transactions from concurrently loading the same data over the network.
% Then, it sends the request to the memory server to access the remote data.
% Once the data is returned, it is cached in the compute server for future accesses, and the latch is released.

% The memory server returns the data to the compute server, which will be cached in local memory for future accesses.
% Upon a cache hit, the compute server accesses the data directly from the cache without involving the memory server.
For a write request, the compute server directly updates the cached data, marks it as dirty, and defers write-back to the memory server until eviction.
% Since it logically owns the partition, the compute server can perform this operation without coordination with other servers.
During eviction of a dirty entry, the data is latched to ensure correctness and prevent concurrent modification during write-back.
Once the updated data is flushed to the memory server, the entry is removed from the cache.
Clean entries are evicted immediately without additional operations.

\begin{figure*}[!t]
    \centering
    \begin{minipage}{0.32\textwidth}
        \centering
        \includegraphics[width=\textwidth]{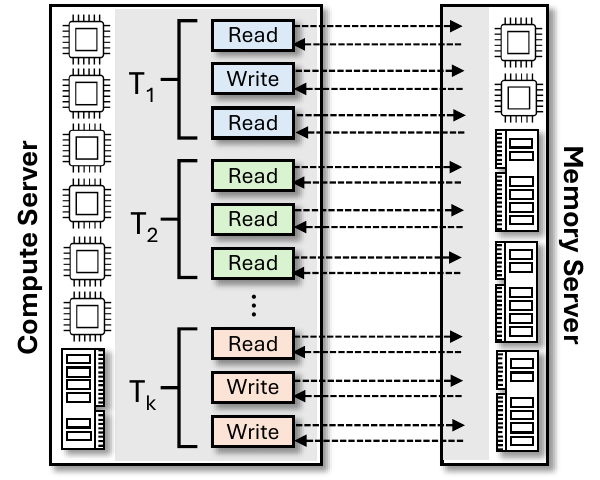}\vspace{-.1in}
        \subcaption{Non-batching.}\vspace{-.15in}
        \label{fig:non_batching}
    \end{minipage}
    \hfill
    \centering
    \begin{minipage}{0.32\textwidth}
        \centering
        \includegraphics[width=\textwidth]{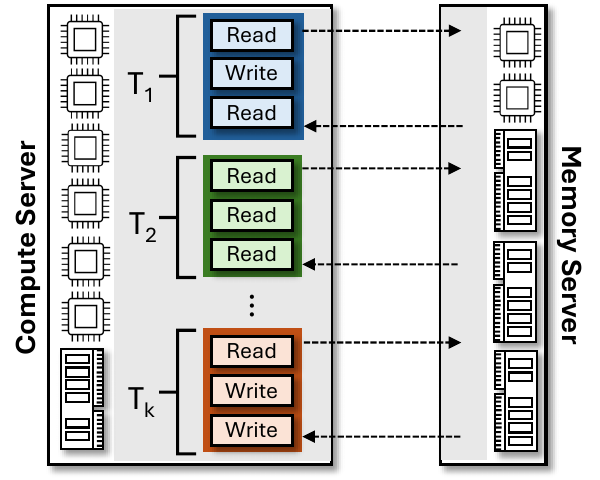}\vspace{-.1in}
        \subcaption{Intra-transaction batching.}\vspace{-.15in}
        \label{fig:intra_batching}
    \end{minipage}
    \hfill
    \centering
    \begin{minipage}{0.32\textwidth}
        \centering
        \includegraphics[width=\textwidth]{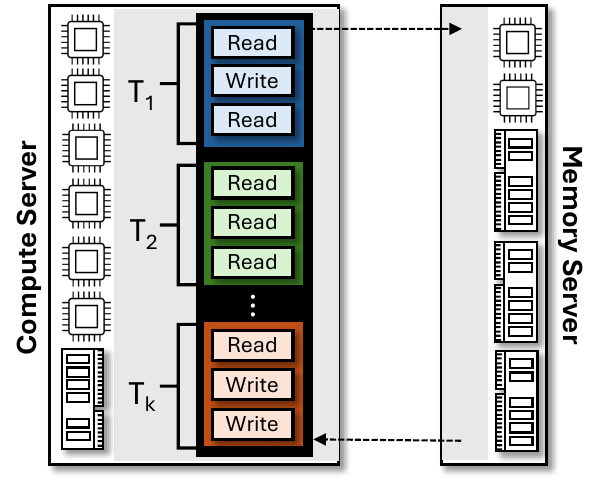}\vspace{-.1in}
        \subcaption{Inter-transaction batching.}\vspace{-.15in}
        \label{fig:inter_batching}
    \end{minipage}
    \caption{Batching techniques in \work.}\vspace{-.1in}
    \label{fig:batching_scenario}
\end{figure*}

\section{Batching}
\label{sec:batch}

In this section, we present our batching techniques, a key optimization that fundamentally reshapes the performance trade-offs of two-sided RDMA by substantially reducing network round-trips between compute servers and memory servers.
% for two-sided RDMA, that can efficiently reduce the number of network round-trips between compute servers and memory servers.

While prior studies have evaluated two-sided RDMA against one-sided RDMA in various distributed systems~\cite{namdb, pilaf, drtm},
they mainly focus on workloads for key-value stores or RPC systems that handle a single request at a time.
Figure~\ref{fig:non_batching} illustrates such cases, where each request in a transaction incurs a separate network round-trip.
However, this per-request communication can be inefficient, particularly in disaggregated memory systems, where many requests require remote memory accesses.
% However, this per-request communication is suboptimal, particularly in disaggregated memory systems, where many requests require remote memory accesses.

We make a key observation that database transactions usually consist of multiple requests that can be issued and processed concurrently.
We exploit this property by aggregating multiple requests targeting the same memory server into a single two-sided RDMA message.
This optimization is inherently specific to two-sided RDMA, as its simple request-response communication abstraction naturally supports aggregation of requests and responses, whereas one-sided RDMA decomposes each request into multiple complex network operations for indexing and concurrency control.
Leveraging this insight, we introduce two batching techniques: (1) intra-transaction batching and (2) inter-transaction batching.
% We exploit this property to batch multiple requests, \hcha{that fall into the same memory server,} into a single message.
% This is a unique optimization technique for two-sided RDMA, as processing a request with one-sided RDMA requires multiple operations to process indexing and concurrency control.
% In the following, we present two batching techniques: (1) intra-transaction batching and (2) inter-transaction batching.

\subsection{Intra-Transaction Batching}
\label{sec:intra_batching}
Intra-transaction batching combines multiple requests within a transaction into one network message.
Figure~\ref{fig:intra_batching} shows an example of intra-transaction batching.
When a compute server executes a transaction, it first collects multiple requests that can be processed concurrently.
Then, it sends those requests to a memory server via an RDMA SEND operation.
Upon receiving the batched requests, the memory server processes them sequentially and returns their results to the compute server with another RDMA SEND.
% \yxy{does the memory server initiate the RDMA SEND? I thought only compute servers can initiate RDMA SEND}\hcha{compute servers initiate it. it posts RDMA SEND and then poll its QP to check if received a response which is basically RDMA RECV. In the memory server side, after receiving the request from RDMA RECV, it posts RDMA SEND to return the result. So it's always 1-to-1 matching operation. Modified}
This approach minimizes the number of network round-trips per transaction, significantly reducing network processing overhead.
% \yxy{One small complication: you can batch only requests that go to the same memory server. Maybe worth mentioning somewhere.}\hcha{Mentioned it in the previous paragraph.}

While batching opportunities vary across transactions, some may have few requests and may include dependencies, still requiring multiple round-trips.
To further improve batching effectiveness in such cases, we additionally apply inter-transaction batching.
% However, this batching may not be always effective.
% Some transactions may only have a small number of requests, and some may consist of requests that have dependencies.
% Those transactions still require multiple network round-trips, which cannot exhibit sufficient batching effects.
% In such cases, we further apply inter-transaction batching to maximize the batching efficiency.

\subsection{Inter-Transaction Batching}
\label{sec:inter_batching}
Inter-transaction batching aggregates requests from multiple transactions into a single network message to further reduce network round-trips.
Figure~\ref{fig:inter_batching} illustrates an example of inter-transaction batching.
In each compute server, \work organizes threads that execute transactions concurrently into several batch groups (e.g., 8 threads per group).
Instead of sending requests individually as in intra-transaction batching, group members submit their requests to a group leader.
The leader then collects and sends these requests to a memory server with a single RDMA SEND.
% Then, the leader collects those requests for a certain period of time and sends them to a memory server with a single RDMA SEND/RECV operation.
The memory server processes the batch of requests and returns the results to the leader with another RDMA SEND.
Finally, the leader dispatches the results to the members.
Next, we discuss its design details.

% By amortizing the communication overhead across multiple transactions, inter-transaction batching can significantly reduce the total number of network round-trips.
% % This batching technique can further reduce the number of network round-trips significantly.
% In the following, we discuss the design details to fully leverage its performance potential.
% However, it also introduces new challenges that may degrade the batching efficiency.
% In the following, we discuss those challenges and present our solutions.
% \yxy{try not to say these as challenges. Maybe something like: the following design details/considerations to fully leverage its performance potential.}\hcha{Modified it.}

\subsubsection{Leader Selection}
Selecting an appropriate group leader is critical for efficient batching.
A static leader may delay batching progress because
% A static group leader may cause the group members to wait for a long time.
if the leader's transaction aborts and enters backoff, other members must wait until the leader comes back.
% For example, if a member, that is the leader of a group, aborts its transaction, it may take a backoff before the transaction re-execution.
% In the meantime, the other members, that have submitted their requests, have to wait for the leader to complete the backoff and collect the requests from the group members.

\work addresses this issue by employing dynamic leader selection.
Each group maintains a leader ID. %that indicates whether a leader is currently active.
When a member finishes submitting its requests, it checks whether a leader exists.
% When a member finishes submitting its requests, it checks whether a leader exists in the group.
If no leader is present, the member atomically updates the leader ID to its own ID and becomes the new leader.
Then, the leader begins collecting requests from other group members.
% becomes the new leader by atomically updating the leader ID variable with its thread identifier and begins collecting requests from the other members.
% If a leader in the group has not been selected, it becomes the new leader and collects requests from the other members.
The leader serves for multiple batches to reduce the coordination overhead of leader selection.
Its term ends when the batch window expires or its transaction aborts.
The leader then resigns by resetting the value of the leader ID, and the next member that updates it becomes the new leader.
% Upon resignation, the first member who has submitted requests subsequently takes over the leadership.

\subsubsection{Batch Formation}
Timely batch formation is essential to balance batching efficiency and request latency.
% Similarly, delayed batch formation may cause additional latency.
% Transactions that submit requests too early or too late may otherwise experience unnecessary waiting.
% Some transactions may have to wait for a long time if they submitted their requests too early or too late.
\work employs a timeout-based fallback mechanism with dynamic waiting times.
A batch manager maintains lightweight statistical metrics for each member such as batch success rate, timeout rate, and conflict rate.
For each batching decision, it computes an appropriate timeout value based on these metrics to avoid excessive delays.
If a member's request is not claimed by the leader within the waiting time, it falls back to individual request transmission.
Note that these statistics do not require coordination among group members.
% we set timeout thresholds and dynamically adjust a member's waiting time based on recent transaction patterns.
% That is, the batch manager keeps track of statistical metrics for each group member such as the batch success rate, timeout rate, and conflict rate.
% For each batching decision, the manager calculates the optimal timeout value based on these metrics, to avoid excessive waiting time.
% Note that these statistics are easy to collect and process, and does not require any coordination between group members.
% \yxy{maybe mention that such statiscs are easy to collect and process}\hcha{Added a sentence mentioning it.}

\subsubsection{Contention-awareness}
Workload contention may affect batching effectiveness.
Under high contention, aborted transactions may temporarily leave the group and rejoin after backoff, reducing the group utilization.
As a result, the leader may collect fewer requests, and members may time out and fall back to individual transmission.
% Consequently, the leader may collect fewer requests, and members may fall back to sending requests individually after timing out.
% High contention workloads may degrade the batching efficiency.
% For example, aborted transactions will leave the group and then re-join it after a backoff.
% Therefore, the group leader may not be able to collect enough number of requests within a certain period of time.
% Moreover, the members that have already submitted their requests may timeout and decide to send their requests individually, which further hurts the batching efficiency.

\work addresses this issue in two ways.
First, each group maintains the number of active members and adjusts batching decisions accordingly.
When the count falls below a threshold, 
% requests are sent individually rather than waiting for batch formation
members send requests individually instead of waiting for batch formation.
% If the number of active members is below a certain threshold, the group members decide to send their requests individually instead of waiting for batch formation.
% Second, statistical metrics guide batching decisions dynamically.
Second, \work uses the statistical metrics to guide batching decisions dynamically.
% We also leverage the statistical metrics to dynamically make batching decisions.
Rather than batching all requests indiscriminately, each group selectively applies batching based on the observed patterns.
% Rather than batching all requests indiscriminately, each group selectively applies batching based on observed patterns, allowing members to individually transmit requests when it is more beneficial.
% When a member finds that sending individual requests is more beneficial than batching, it sends the requests individually.

\subsubsection{Adaptivity}
% Finally, these parameters for dynamic batching may not be effective under varying workloads, and may lead to making suboptimal decisions.
% \yxy{saying a design "may not be effective" sounds negative, instead, you can say "in order to improve effectiveness under varying workloads, ..." which sounds more positive}\hcha{you're right. revised it.}
In order to improve effectiveness under varying workloads, \work maintains batching parameters in an adaptive manner.
% that learns from the workload patterns.
For each network request, the system records statistics that capture recent batching behavior,
and for each batch window, these parameters are updated using an exponentially weighted moving average (EWMA)~\cite{ewma}, enabling the batching mechanism to smoothly respond to workload changes over time.
% For each network request, the system records statistics to reflect the recent batching patterns,
% and, for each batch window, we maintain these parameters with exponentially weighted moving average (EWMA)~\cite{ewma} to adapt to changing workloads.

\subsection{Discussion}
\label{sec:batch_discussion}

Prior work has proposed doorbell batching to accelerate two-sided RDMA operations~\cite{herd, fasst}.
This technique reduces the overhead of memory-mapped I/Os by combining multiple RDMA work requests into a single doorbell at the network level.
However, it does not reduce the number of network round-trips and is typically applicable to systems where multiple threads share a single queue pair, a design commonly used with unreliable datagram (UD) transport, which does not guarantee reliable message delivery.
% To handle packet losses in UD, prior work has applied a timeout-based mechanism that restarts the entire application when a packet loss is detected~\cite{fasst}.

% Another critical limitation of UD is that it does not support messages larger than the maximum transfer unit (MTU) of the hardware (e.g., 4~KB on Mellanox ConnectX-5 NICs).
% This imposes a strict limit on packet sizes and requires an application-level mechanism for packet fragmentation at the sender and reassembly at the receiver, which causes additional network round-trips.  

In contrast, \work presents lightweight batching techniques at the software level.
Unlike doorbell batching, our approaches combine multiple transaction requests into a single RDMA operation, thereby reducing the number of network round-trips.
These techniques are generally applicable across distributed systems regardless of the underlying network transport, including both RDMA and traditional socket-based networks.
\work supports both UD and reliable connection (RC) and uses RC by default to ensure reliable message delivery while supporting varying message sizes.
% , particularly for our batching techniques.

Although reliability can be provided over UD, it comes with limitations.
For example, prior work uses a timeout-based mechanism to detect and handle packet loss in UD~\cite{fasst}.
However, this requires restarting the entire system upon a loss event.
Moreover, UD imposes a strict packet size limit, since it does not support packets larger than the hardware's maximum transfer unit (MTU), typically only a few KB~\cite{nvidia_rdma_userguide}.
% Moreover, UD imposes a strict limit on packet size since it does not support messages larger than the hardware's maximum transfer unit (MTU), (e.g., 4~KB on Mellanox ConnectX-5 NICs)~\cite{nvidia_rdma_userguide}.
This requires application-level mechanisms for packet fragmentation at the sender and reassembly at the receiver, which adds complexity and incurs additional network round-trips.

We investigate the performance of RC and UD in Section~\ref{sec:rc_ud}.
\section{Evaluation}
\label{sec:evaluation}

% This section evaluates \work.
We divide our performance study into two parts based on the memory disaggregation models, non-partitioned memory and co-partitioned memory.
% , because they exhibit different behaviors in indexing and concurrency control, as discussed in Section~\ref{sec:motivation}.
We first describe our experimental setup and benchmarking workload in Sections~\ref{sec:setup} and \ref{sec:workload}, respectively.
We outline the details of our benchmarking system, indexing, and concurrency control baselines in Section~\ref{sec:implementation}.
Then, we present and analyze the performance of \work from Sections~\ref{sec:scalability} to \ref{sec:mem_threads_sensitivity}.
% \yxy{would be nice to show a short bullet list, highlighting what the experiments aim to show (e.g., what Lotus achieves through 2-sided design), like more sophisticated CC, improved perf through caching, and improved perf through batching, etc.} \hcha{Added bullets points below.}
In particular, we aim to answer the following questions in our evaluation:%\vspace{-.03in}
\begin{itemize}[leftmargin=*]
    \item How sophisticated concurrency control in \work achieves starvation freedom and sustains high performance under contention.
    \item How caching in \work improves performance by reducing remote memory accesses.
    \item How batching techniques in \work reduce network round-trips and improve throughput.
\end{itemize}

\subsection{Experimental Setup}
\label{sec:setup}

We conduct all experiments on four CloudLab~\cite{cloudlab} machines (c6525-100g instances).
Each machine contains an AMD EPYC 7402P CPU (24 cores/48 hyper-threads), 128~GB of DRAM, and a 100~Gbps Mellanox ConnectX-5 NIC.
Following prior work on RDMA-based disaggregated memory systems~\cite{sherman, dex, deft}, we configure each machine to act as a compute server and a memory server.
On each machine, we allocate 40 threads for the compute server and 8 threads for the memory server, following configurations in prior work~\cite{namdb, namtree}.
We also evaluate sensitivity to the number of memory server threads in Section~\ref{sec:mem_threads_sensitivity}.
% To avoid unnecessary CPU migration, we pin threads to specific cores for each server.
Similar to a recent study~\cite{dex}, each compute server is provisioned with a 128~MB cache (i.e., 8\% of the index data size).
% Similar to a recent study~\cite{dex}, we allocate 128~MB of DRAM cache in each compute server (i.e., 8\% of the index data size) in our evaluation.
After loading the data, we warm up the DBMS and then measure the performance by running each workload for 10 seconds.
% \yxy{what is the fault tolerant aspect of the story? Even if we have no such support we should mention it somewhere.} \hcha{Added discussion about fault tolerance in 7.2.4}

\subsection{Workload}
\label{sec:workload}

% We use two different types of benchmark workloads: (1) YCSB, and (2) TPC-C for the evaluation.
We use Yahoo! Cloud Serving Benchmark (YCSB)~\cite{ycsb}, a widely used benchmark for key-value store evaluation.
% \noindent\textbf{YCSB.}
% The Yahoo! Cloud Serving Benchmark (YCSB)~\cite{ycsb} is a widely-used key-value store benchmark.
We use a 100~GB database scale consisting of a single table with 100 million records.
Each record is 1~KB in size, containing a single primary key and 10 additional columns of randomly generated string data.
% The DBMS maintains a B+-tree index for the primary key.

Each transaction accesses 16 records by primary key.
Record accesses follow a Zipfian distribution ($\theta=0.9$), and each access is either a read or write.
% We use three workload configurations, \texttt{A}, \texttt{B}, and \texttt{C}, to capture different levels of read-write contention.
% Workload \texttt{A} is write-intensive, with 50\% reads and 50\% writes. Workload \texttt{B} is read-mostly, with 95\% reads and 5\% writes. Workload \texttt{C} is read-only.
We use three workload configurations to capture different levels of read-write conflicts:
% We use three workload configurations to provide a comprehensive evaluation under different read-write conflict levels:
\begin{itemize}[leftmargin=*]
    \item Workload \texttt{A} is write-intensive, with 50\% reads and 50\% writes.
    \item Workload \texttt{B} is read-mostly, with 95\% reads and 5\% writes.
    \item Workload \texttt{C} is read-only.
\end{itemize}
% Workload \texttt{A} is write intensive, with 50\% reads and 50\% writes.
% Workload \texttt{B} is read mostly, with 95\% reads and 5\% writes.
% Workload \texttt{C} is read only.

% \noindent\textbf{TPC-C.}
% TPC-C benchmark~\cite{tpcc} is a representative OLTP benchmark.
% It simulates a warehouse-centric order processing application, and each warehouse consists of nine tables.
% We use three database size settings, 4, 16, and 64 warehouses, to evaluate under different levels of contention.
% We follow the standard TPC-C specification for the database schema and transaction mix ratio~\cite{tpcc}.

\subsection{Implementation}
\label{sec:implementation}

\subsubsection{Testbed}
% \noindent\textbf{Testbed.} 
We implement 
% \work\footnote{Source code of \work is available at https://anonymous.4open.science/r/Lotus-51F4} 
on top of DBx1000~\cite{dbx1000}, an in-memory DBMS research prototype.
We extend it to a disaggregated system with RDMA by separating the computing layer and the memory layer of the DBMS.
% We also replace the hash table indexes with B+-tree indexes.

\subsubsection{Index}
% \noindent\textbf{Index.}
The DBMS maintains a primary index for each table, where each index stores pointers to data records.
For \work, we use a B+-tree that employs optimistic latch coupling for its synchronization~\cite{olfit2001}.
We compare \work with two state-of-the-art RDMA-based B+-tree indexes, i.e., Sherman/Deft~\cite{sherman, deft} and DEX~\cite{dex}, which all use one-sided RDMA for remote memory access.
\begin{itemize}[leftmargin=*]
    \item \textbf{Sherman}~\cite{sherman} is a write-optimized B+-tree designed for non-partitioned memory that leverages shortcut hint caching.
    \item \textbf{Deft}~\cite{deft} extends Sherman by segmenting tree nodes to reduce network amplification and uses shared-exclusive latching.
    \item \textbf{DEX}~\cite{dex} targets co-partitioned memory to eliminate cache coherence across compute servers and employs opportunistic operation pushdown to memory servers.
\end{itemize}

We use the open-source implementations of Sherman and DEX from the authors.
We use the same parameter settings as in prior work, e.g., RPC pushdown rate in DEX.
We modify Sherman's optimistic synchronization since its original version may lead to incorrect behavior~\cite{rdma_guideline}.
We do not directly use Deft because it heavily relies on experimental one-sided RDMA atomic operations, i.e., extended CAS and extended masked FAA, which were deprecated in 2020\footnote{https://docs.nvidia.com/networking/display/ofedv512580/release+notes}.
Instead, we adjust the node size in Sherman to match the segmented node size in Deft to demonstrate its reduced network I/O.

\subsubsection{Concurrency Control}
% \noindent\textbf{Concurrency Control.}
We evaluate three 2PL protocols: \nowait, \waitdie, and \woundwait~\cite{2pl1979, 2pl1981}.
% We focus on three concurrency control protocols in 2PL: \nowait, \waitdie, and \woundwait~\cite{2pl1979, 2pl1981}.
For \work, we implement all three protocols using two-sided RDMA, as discussed in Section~\ref{sec:concurrency_control}, together with two-phase commit (2PC)~\cite{2pc}.
For Sherman/Deft, we implement \nowait and \waitdie using one-sided RDMA, following prior work~\cite{rcc}.
\woundwait is not included because implementing preemption is extremely difficult due to limited primitives of one-sided RDMA, as discussed in Section~\ref{sec:rdma_txn}.
Note that \waitdie in Sherman/Deft does not guarantee starvation prevention for the same reason.
For co-partitioned memory systems, i.e., \work and DEX, we focus on indexing performance.
As discussed in Section~\ref{sec:rdma_cc}, concurrency control in co-partitioned memory can be handled locally in compute servers, thus the system performance depends solely on indexing efficiency.

% One possible way to reduce latency is to place another 8-byte value next to the lock to maintain the priority of the current lock owner~\cite{rcc}.
% After lock acquisition, the owner updates the priority value of the lock with its transaction priority via RDMA CAS, letting other transactions determine to wait or abort upon a conflict.
% However, this still does not provide starvation prevention.
% % When a lock is released, its ownership is not guaranteed to be handed over to the next waiter with the highest priority.
% Since the lock should be acquired and released via RDMA CAS, whichever transaction that succeeds to swap the lock value becomes the next owner regardless of their priorities.

\subsubsection{Fault Tolerance}
While fault tolerance is a critical component in distributed systems, disaggregated architectures still present many open design choices for implementing logging and recovery, particularly with respect to network transports.
Some systems adopt a conservative approach by employing centralized log servers~\cite{aurora, socrates, polardb, fusee}, while others decentralize logging by decomposing conventional fault tolerance protocols to independently handle failures in compute servers and memory servers~\cite{ddpm, legobase, polardbMP}.
Moreover, these design choices fuel an ongoing debate between one-sided RDMA and two-sided RDMA for logging and recovery~\cite{gam, active_memory, tailwind, dsmdb}.
Therefore, we leave fault tolerance as future work and focus on indexing and concurrency control in this paper.

\subsection{Scalability Analysis}
\label{sec:scalability}

% \yxy{font size a bit small in all figures.} \hcha{Made fonts larger in all figures.}
\begin{figure}[!t]
    \centering
    \includegraphics[width=0.47\textwidth]{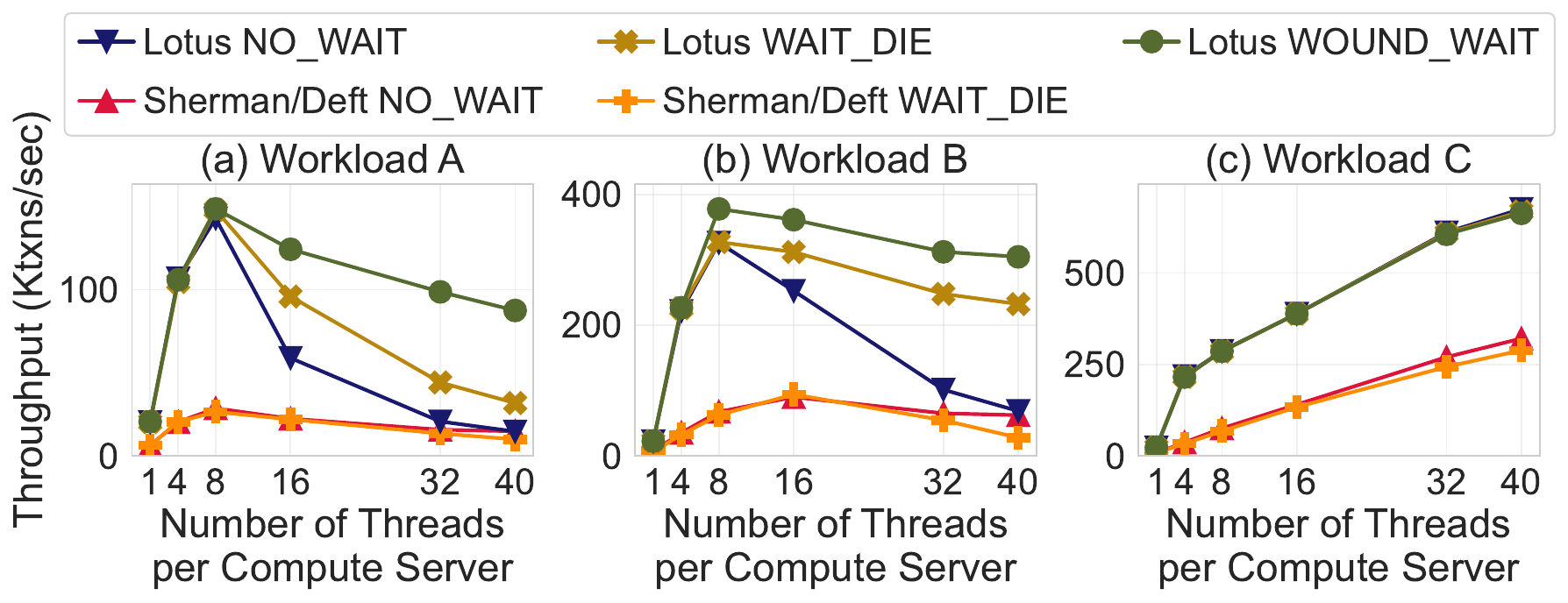}\vspace{-.1in}
    \caption{Throughput with a varying number of compute threads in YCSB workloads.}\vspace{-.15in}
    \label{fig:ycsb_general_throughput}
\end{figure}

\subsubsection{Throughput}
We first evaluate the scalability of \work against Sherman/Deft.
Figure~\ref{fig:ycsb_general_throughput} shows the throughput for workloads \texttt{A}, \texttt{B}, and \texttt{C} with a varying number of compute threads.

For workloads \texttt{A} and \texttt{B}, all schemes do not scale due to conflicts.
However, \work shows much higher throughput than Sherman/Deft.
Sherman/Deft lacks functional capabilities to implement starvation prevention and priority-based scheduling due to limited APIs of one-sided RDMA.
This significantly degrades performance when frequent read-write conflicts occur.
% In particular, \work shows much higher throughput than Sherman/Deft.
The impact of high contention on \work is much smaller, because of its rich functionalities in concurrency control that handle conflicts based on transaction priorities.
Specifically, \work \woundwait achieves the highest throughput and sustains good performance with an increasing number of compute threads as it avoids excessive lock thrashing.

% Sherman/Deft lacks functional capabilities to implement starvation prevention and priority-based scheduling in concurrency control due to limited APIs of one-sided RDMA.
% This significantly degrades the performance when frequent read-write conflicts occur.
% % This causes significant performance degradation, especially in this kind of workloads, where frequent read-write conflicts occur.
% In contrast, \work effectively handles the conflicts based on transaction priorities with rich functionalities in concurrency control.
% Specifically, \work \woundwait achieves the highest throughput and maintains good performance with an increasing number of compute threads, as it avoids excessive lock thrashing.
% \waitdie schemes show higher throughput than \nowait schemes with priority-based scheduling.

For workload \texttt{C}, all schemes scale well as the number of compute threads increases, since there are no read-write conflicts.
\work generally outperforms Sherman/Deft.
In Sherman/Deft, network amplification limits system performance since each transaction request requires multiple RDMA READs for index traversal.
In contrast, \work minimizes the number of network round-trips by batching multiple requests in a network message in two ways, i.e., intra-transaction and inter-transaction.

% \begin{figure}[!t]
%     \centering
%     \includegraphics[width=0.48\textwidth]{figures/ycsb_general_latency.pdf}
%     \caption{CDF of latency in YCSB workloads, 40 threads per compute server.}
%     \label{fig:ycsb_general_latency}
% \end{figure}

\begin{figure}[!t]
    \centering
    \includegraphics[width=0.47\textwidth]{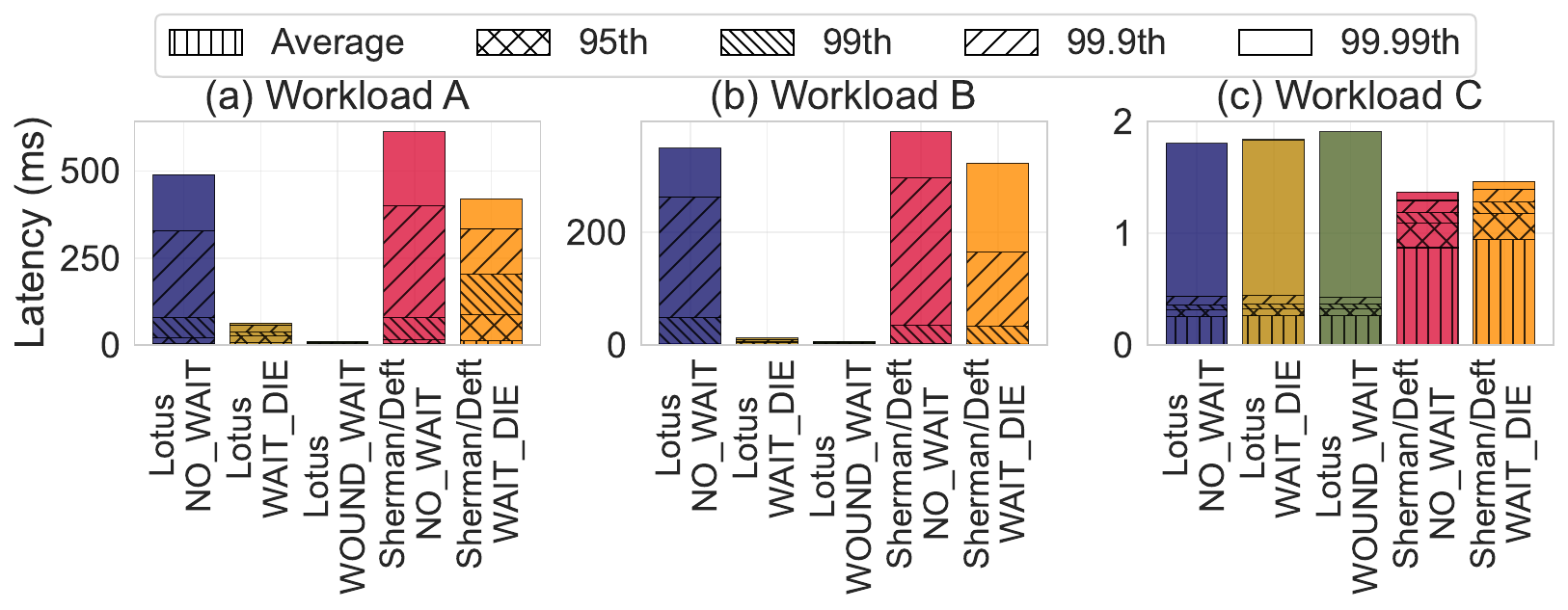}\vspace{-.15in}
    \caption{Latency percentile breakdown in YCSB workloads, 40 threads per compute server.}\vspace{-.15in}
    \label{fig:ycsb_general_latency_bar}
\end{figure}

\subsubsection{Latency}
% \yxy{It might help to show another figure that highlights tail latency. Right now it's a bit hard to see from Figure 6. One option is to report a bar graph with 99th or 99.9th percentile latency. Also, you might want to mention that NO\_WAIT actually has a lower average latency, but a bad tail latency hurts its throughput. } \hcha{Added the bar graph, and separated scalability subsection into throughput and latency subsubsections. We might want to remove one of the two figures (CDF / bar) if it is too redundant, but they have different trade-offs...}

Next, we analyze the latency of \work and Sherman/Deft. 
Figure~\ref{fig:ycsb_general_latency_bar} shows the latency percentile breakdown.
% Figure~\ref{fig:ycsb_general_latency_bar} shows the CDF of latency for YCSB workloads.
% We also provide latency percentile breakdown in Figure~\ref{fig:ycsb_general_latency_bar} for clarity.

For workloads \texttt{A} and \texttt{B}, \nowait schemes show high tail latency as they do not consider transaction priorities upon conflicts.
While \waitdie schemes report lower latency,
the latency of Sherman/Deft \waitdie is much higher than \work \waitdie since Sherman/Deft \waitdie does not guarantee starvation prevention, as discussed in Section~\ref{sec:rdma_cc}.
\work \woundwait shows the lowest latency as it allows preemption in addition to priority-based scheduling.

For workload \texttt{C}, all schemes report much lower latency than workloads \texttt{A} and \texttt{B} as it is read-only.
\work shows lower average latency than Sherman/Deft because \work minimizes the number of network round-trips with batching, while Sherman/Deft suffers from network amplification.
However, \work shows higher tail latency due to coordination overhead in inter-transaction batching.

\subsection{Network Traffic Analysis}
\label{sec:network_traffic}

\begin{figure}[!t]
    \centering
    \includegraphics[width=0.48\textwidth]{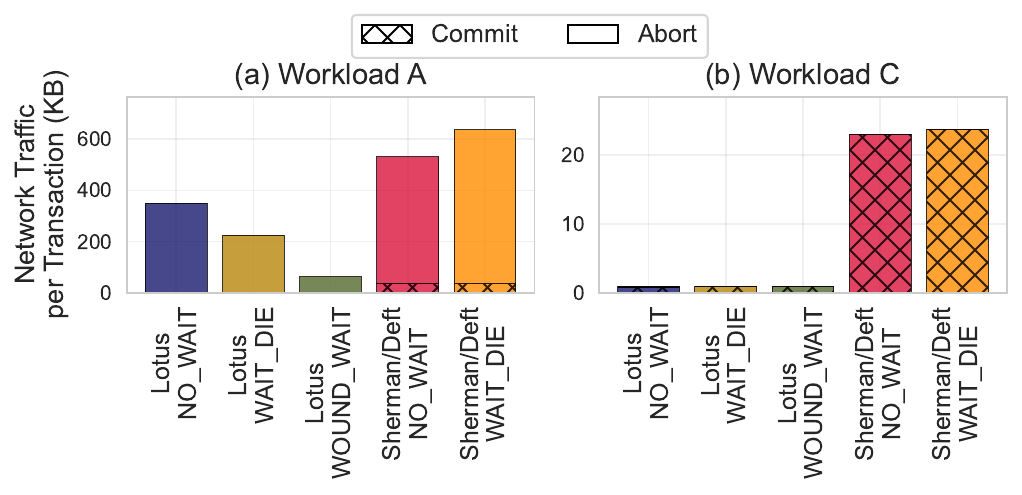}\vspace{-.15in}
    \caption{Network traffic in YCSB-\texttt{A} and \texttt{C}.}\vspace{-.15in}
    %\yxy{is it possible to add some traffic breakdown? like request, response, abort (traffic from a txn that aborted), etc.} \hcha{I did it only for Commit and Abort since the Commit traffic is already small enough in workload A.}
    % }
    \label{fig:ycsb_network_traffic}
\end{figure}

To understand the impact of network amplification, we measure the network traffic of each scheme during the execution of YCSB workloads. 
For \work, we measure the traffic of two-sided RDMA operations, i.e., SEND and RECV. For Sherman/Deft, we measure the traffic of one-sided RDMA operations, i.e., READ, WRITE, CAS, and FAA.
We do not include the traffic of data transfers for committed transactions, as all schemes transfer the same amount.

Figure~\ref{fig:ycsb_network_traffic} shows the average network traffic per transaction for workloads \texttt{A} and \texttt{C}.
All schemes generate more traffic in workload \texttt{A} than in \texttt{C}, due to read-write conflicts.
However, \work consistently incurs much less traffic than Sherman/Deft.
% While all schemes generate more traffic in workload \texttt{A} than in workload \texttt{C} due to read-write conflicts, \work generally transfers much less data compared to Sherman/Deft.
This is because \work only transfers request and response information in network messages.
Although Sherman/Deft employs caching, it still needs to traverse the index from the root node to a leaf node upon a cache miss and further requires multiple RDMA READs for correct synchronization during traversal.

In workload \texttt{A}, \work \waitdie and \work \woundwait generate less traffic compared to \work \nowait since priority-based scheduling and preemption effectively handle conflicts, reducing the amount of transaction aborts.
However, Sherman/Deft \waitdie incurs more traffic than Sherman/Deft \nowait due to additional priority management.
Sherman/Deft \waitdie requires extra RDMA operations to check and update transaction priorities in lock acquisition.
In workload \texttt{C}, \work schemes show significantly reduced network traffic compared to Sherman/Deft schemes by minimizing the number of network round-trips via batching.

\subsection{Varying Transaction Lengths}
\label{sec:ycsb_txn_length}

\begin{figure}[!t]
    \centering
    \includegraphics[width=0.47\textwidth]{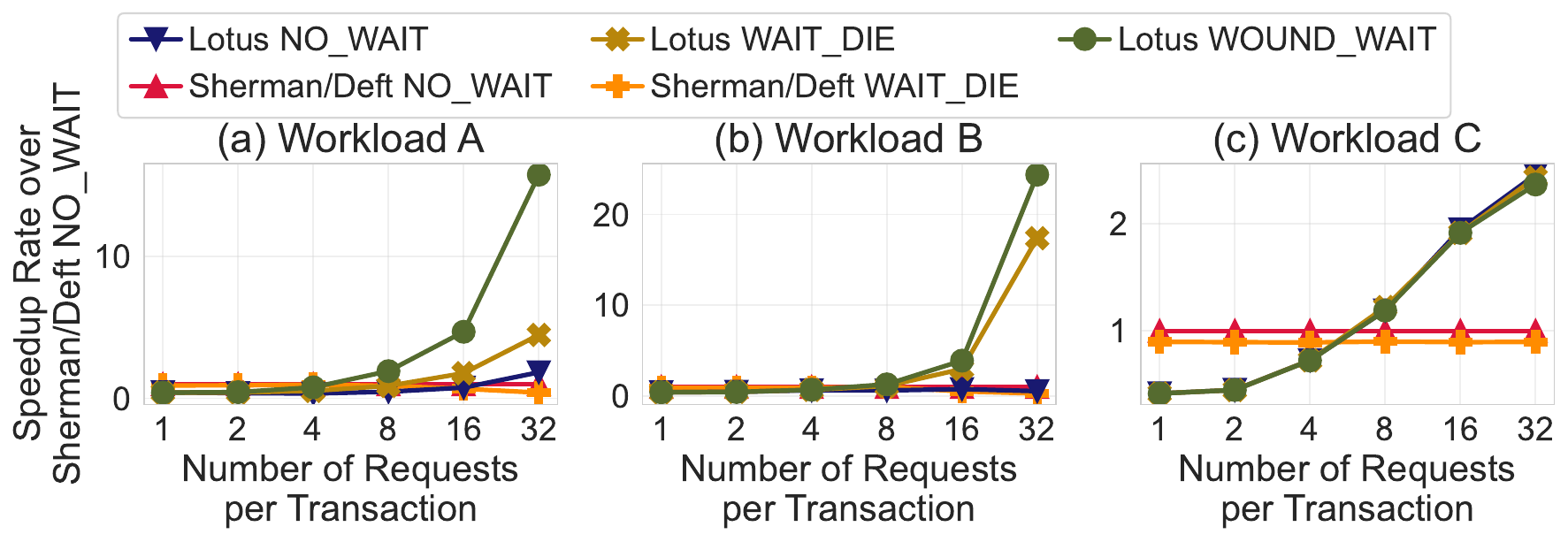}\vspace{-.15in}
    \caption{Relative performance with varying transaction lengths in YCSB workloads, 40 threads per compute server.}\vspace{-.1in}
    \label{fig:ycsb_requests_throughput}
\end{figure}

\begin{table}[!t] 
    \centering
    \caption{Average number of requests per message in \work.}\vspace{-.15in}
    \label{tab:requests_per_message}
    {\small
    \setlength{\tabcolsep}{6pt}
    \renewcommand{\arraystretch}{0.5} % reduce row height
    \begin{tabular}{c | c | c | c | c | c | c}
        \toprule
        \diagbox[width=8em, innerleftsep=1pt, innerrightsep=1pt, dir=NW]
        {\makebox[0pt][l]{\footnotesize\textbf{Workload}}}
        {\makebox[0pt][r]{\footnotesize\textbf{Req/Txn}}}
% \diagbox{\textbf{Wkld}}{\textbf{Req./Txn}}
        % & \multicolumn{6}{c}{\textbf{Req./Txn}} \\
        % \cmidrule(lr){2-7}
        % \textbf{Workload} 
        & \textbf{1} & \textbf{2} & \textbf{4} & \textbf{8} & \textbf{16} & \textbf{32} \\
        \midrule
        \texttt{A} & 1.14 & 1.21 & 1.53 & 2.28 & 4.09 & 8.07 \\
        \texttt{B} & 1.17 & 1.48 & 1.91 & 2.33 & 4.11 & 8.12 \\
        \texttt{C} & 1.18 & 1.53 & 2.25 & 3.82 & 7.16 & 12.77 \\
        % \texttt{A} & 5.84 & 6.67 & 6.89 & 7.91 & 13.45 & 22.42 \\
        % \texttt{B} & 5.95 & 8.91 & 10.11 & 11.91 & 13.65 & 24.31 \\
        % \texttt{C} & 5.97 & 9.11 & 13.38 & 18.73 & 23.08 & 26.24 \\
        \bottomrule
    \end{tabular}
    }
    \vspace{-.1in}
\end{table}

\begin{figure}[!t]
    \centering
    \includegraphics[width=0.47\textwidth]{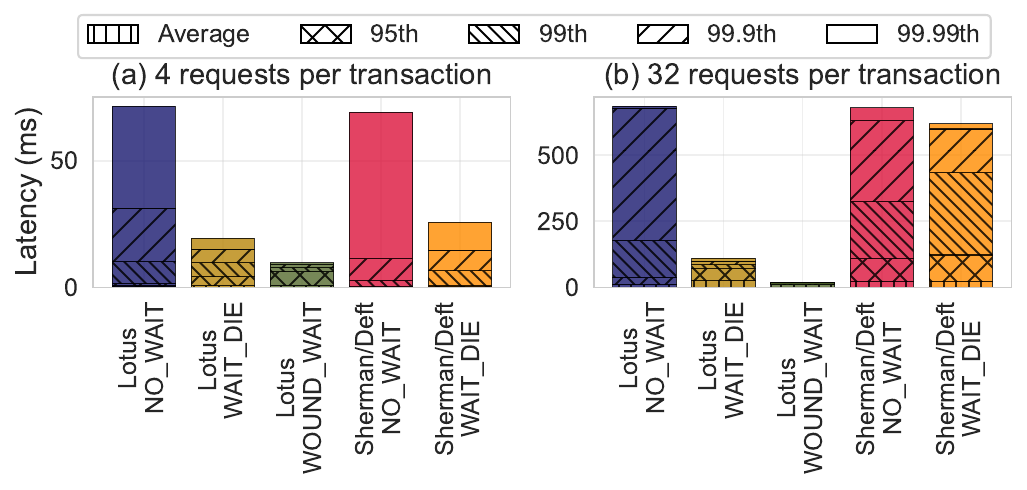}\vspace{-.2in}
    \caption{Latency percentile breakdown in YCSB-\texttt{A}, 40 threads per compute server.}\vspace{-.15in}
    \label{fig:ycsb_requests_latency}
\end{figure}

Next, we examine how transaction length affects performance.
We vary the number of requests per transaction from 1 to 32 to provide a comprehensive performance analysis from short transactions to long transactions.
We report the relative performance of each scheme against Sherman/Deft \nowait since throughput varies substantially across transaction lengths.
% \hcha{I added the last sentence to justify, because one reviewer commented it is unclear why we report relative performance. We might want to remove this throughput figure if we do not have enough space.}

Figure~\ref{fig:ycsb_requests_throughput} shows the relative throughput for YCSB workloads.
Sherman/Deft shows slightly better performance than \work for short transactions, but as the transaction length increases, \work outperforms Sherman/Deft.
\work effectively leverages intra-transaction batching and inter-transaction batching as it can aggregate multiple requests into a network message, which substantially reduces the number of network round-trips.
In contrast, Sherman/Deft handles each request individually using one-sided RDMA, which leads to an increased number of round-trips in proportion to the number of requests in a transaction.
Table~\ref{tab:requests_per_message} reports the average number of requests per message in \work, showing improved batching effects as transactions grow longer.
Note that the averages are lower than the total number of requests per transaction as each transaction spans multiple partitions across memory server, requiring requests to be split into separate messages targeting different servers.

% \begin{figure}[!t]
%     \centering
%     \includegraphics[width=0.48\textwidth]{figures/ycsb_requests_latency.pdf}\vspace{-.05in}
%     \caption{CDF of latency in YCSB-A, 40 threads per compute server.}\vspace{-.1in}
%     % \hcha{I also made a bar graph for this. We can add/replace if necessary.}}
%     % \includegraphics[width=0.48\textwidth]{figures/ycsb_requests_latency_bar.pdf}
%     % \caption{Latency in YCSB-A, 40 threads per compute server.}
%     \label{fig:ycsb_requests_latency}
% \end{figure}

We further analyze the latency of each scheme in workload \texttt{A} under different transaction lengths.
% Figure~\ref{fig:ycsb_requests_latency} shows the CDF of latency for short transactions (4 requests per transaction) and long transactions (32 requests per transaction).
Figure~\ref{fig:ycsb_requests_latency} shows latency percentile breakdown for short transactions (4 requests per transaction) and long transactions (32 requests per transaction).
While all schemes report similar average latency for short transactions, \work \waitdie and \woundwait, as well as Sherman/Deft \waitdie, show lower tail latency due to their consideration of transaction priorities.
For long transactions, the latency gap between protocols that guarantee starvation prevention (i.e., \work \waitdie and \work \woundwait) and those do not (i.e., \work \nowait, Sherman/Deft \nowait, and Sherman/Deft \waitdie) becomes significantly larger.
This highlights the importance of starvation prevention in concurrency control as more frequently transactions abort, they suffer increasingly from lock thrashing.

\begin{figure}[!t]
    \centering
    \includegraphics[width=0.47\textwidth]{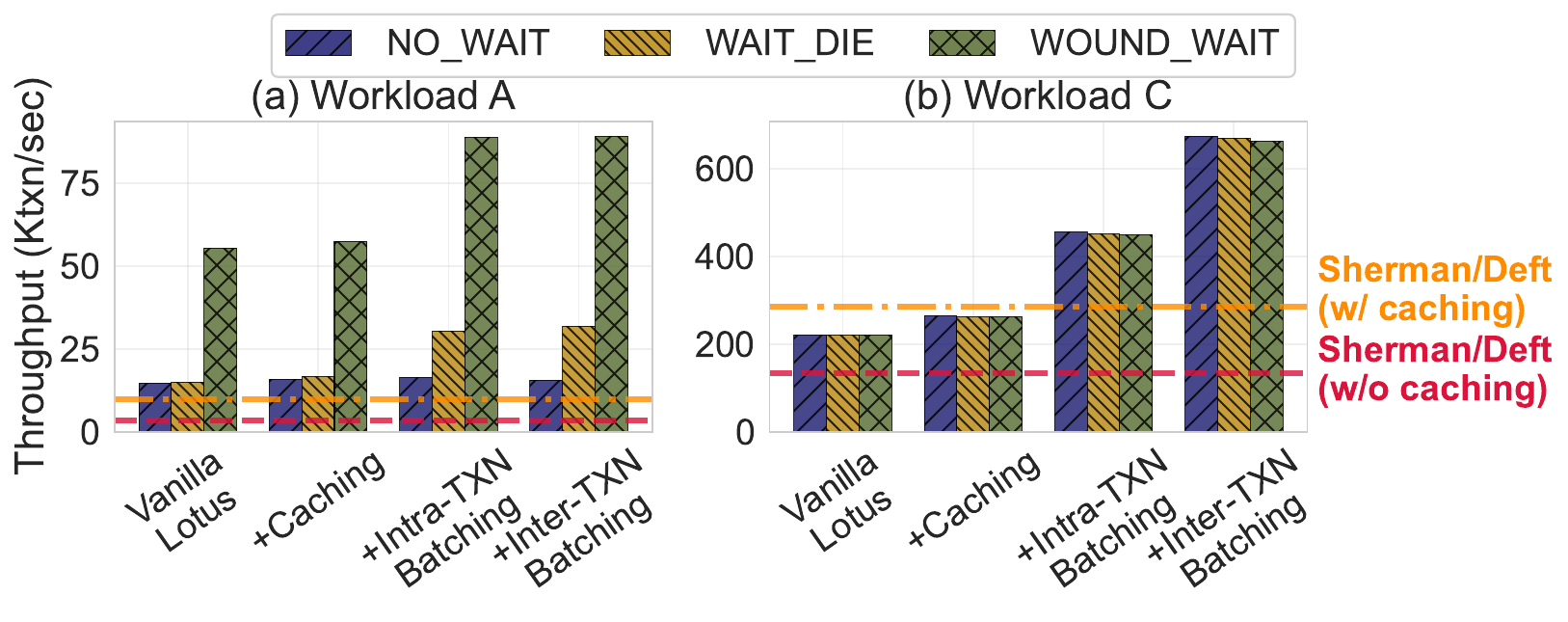}\vspace{-.2in}
    \caption{Throughput analysis of \work in YCSB-\texttt{A} and \texttt{C}.}\vspace{-.15in}
    \label{fig:ycsb_factor_analysis_tput}
\end{figure}

\begin{figure}[!t]
    \centering
    \includegraphics[width=0.47\textwidth]{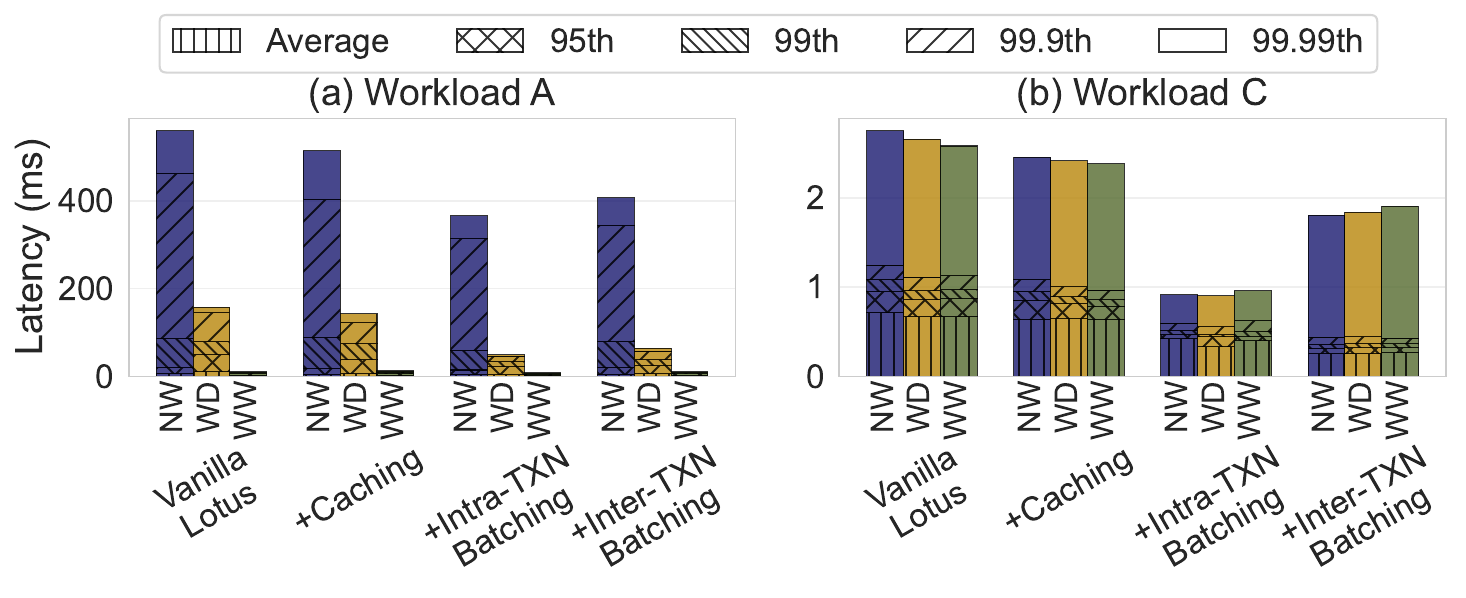}\vspace{-.2in}
    \caption{Latency analysis of \work in YCSB-\texttt{A} and \texttt{C} (NW: \nowait, WD: \waitdie, WW: \woundwait).}\vspace{-.2in}
    % \yxy{In Figure 11(b), is the y-axis also in ms? the y-axis numbers are much smaller than in Fig 11(a).} \hcha{Yes, Fig 11(a) has many read-write conflicts and Fig 11(b) is read-only.}}
    \label{fig:ycsb_factor_analysis_latency}
\end{figure}

\subsection{Analysis of Design Factors}
\label{sec:factor_analysis}

% \yxy{Let's include the baseline in figure 10 as well (as bars or lines), if you can turn off caching in baseline, that's even better, but this is optional. This helps reader to see how each optimization changes the relative performance wrt the baseline.} \hcha{Added Sherman/Deft w/ and w/o caching.}

In this section, we analyze the performance impact of design factors in \work.
We first break down each optimization, i.e., caching, intra-transaction batching, and inter-transaction batching, and apply them incrementally to vanilla \work.

\subsubsection{Throughput and Latency}
Figure~\ref{fig:ycsb_factor_analysis_tput} and \ref{fig:ycsb_factor_analysis_latency} show the throughput and latency of each optimization across concurrency control protocols for YCSB workloads, respectively.
% Figure~\ref{fig:ycsb_factor_analysis} shows the throughput of each optimization across concurrency control protocols in YCSB workloads.
% For comparison, we also report the performance of Sherman/Deft with and without caching as dashed lines.

% \subsubsection{Caching}
% \noindent\textbf{+Caching.}
Caching reduces data processing overhead in memory servers as cache hints help quickly locate data while bypassing index traversal.
This reduces tail latency and improves throughput by up to 1.3$\times$ compared to the baseline, vanilla \work.
% This improves the throughput by up to 1.3$\times$ compared to the baseline, vanilla \work.
% While caching also reduces the latency, its effect is limited since network processing becomes the bottleneck in memory servers as each request requires a network round-trip.
% However, the effect of caching is limited since network processing becomes the bottleneck in memory servers as each request requires a network round-trip.

% \subsubsection{Intra-transaction Batching}
% \noindent\textbf{+Intra-transaction Batching.}
Intra-transaction batching minimizes the number of network round-trips per transaction by combining multiple requests within a transaction into a single network message.
% Batching multiple requests within a transaction into a single network message minimizes the number of network round-trips per transaction.
The throughput improves by up to 1.8$\times$ and 1.7$\times$ for workloads \texttt{A} and \texttt{C}, respectively, compared to the previous increment.
The performance gain is higher in workload \texttt{A} because batching also shortens the duration of locks in proportion to the reduced number of network round-trips, mitigating lock contention.
The reduced number of network round-trips also contributes to significant reduction in tail latency.

% \subsubsection{Inter-transaction Batching}
% \noindent\textbf{+Inter-transaction Batching.}
Inter-transaction batching further reduces the number of network messages by combining requests from multiple transactions in a group into a single message.
% As discussed in Section~\ref{sec:inter_batching}, inter-transaction batching may cause performance degradation in workload \texttt{A}, where read-write conflicts leaf to transaction aborts.
While this significantly improves the performance in workload \texttt{C}, it may cause performance degradation in workload \texttt{A}.
Read-write conflicts lead to transaction aborts, making batch groups difficult to make optimal batching decisions due to backoffs, as discussed in Section~\ref{sec:inter_batching}.
% That is, aborted transactions leave their batch groups and rejoin after backoff, making it difficult for groups to make optimal batching decisions.
The results demonstrate the effectiveness of our adaptive batching strategy, improving the throughput by up to 1.2$\times$ and 1.5$\times$ in workloads \texttt{A} and \texttt{C}, respectively.
% We address this challenge by dynamically adjusting the batching strategy and parameters based on recent statistical patterns so that batch groups can make decisions in an adaptive manner.
% The result demonstrates the effectiveness of our adaptive batching strategy.
% The throughput improves by up to 1.2$\times$ and 1.6$\times$ in workload \texttt{A} and \texttt{C}, respectively.
This also reduces the average latency but increases the tail latency in both workloads due to coordination overhead in batching.
% The tail latency in both workloads increases only marginally, while the average latency in workload \texttt{C} decreases.
% The result demonstrates the effectiveness of our adaptive batching strategy, showing up to 1.2$\times$ and 1.6$\times$ improvement in throughput in workload \texttt{A} and \texttt{C}, respectively.

\subsubsection{CPU Utilization in Memory Servers}
We further analyze the impact of optimization techniques in \work on CPU utilization at memory servers.
For deeper insight, we break down the execution time into five categories:
\begin{itemize}[leftmargin=*]
    \item \textit{Network} includes time spent on RDMA SEND/RECV operations.
    \item \textit{Lock} indicates time spent on concurrency control.
    \item \textit{Index} measures time spent on index traversal.
    \item \textit{Data} includes time spent on data reads and writes.
    \item \textit{Auxiliary} covers system overhead, including message parsing, metadata management, and memory management.
\end{itemize}

\begin{figure}[!t]
    \centering
    \includegraphics[width=0.47\textwidth]{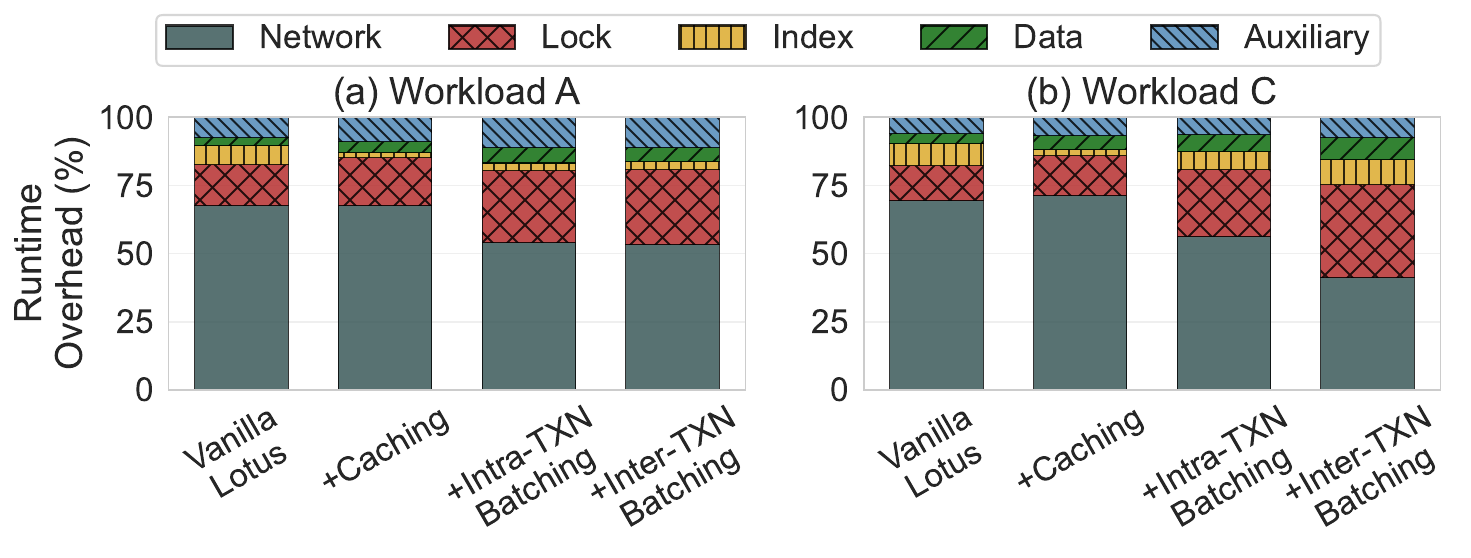}\vspace{-.2in}
    \caption{Runtime breakdown of \work in YCSB-\texttt{A} and \texttt{C}.}\vspace{-.2in}
    % \yxy{I was a bit surprised to see that in workload A, the runtime distribution does not change drastically as we add more optimizations. In figure 10, the throughput looks much more drastically different, and I was expecting a similar visual effect here. For example, I would expect batching to reduce network traffic by several times. Then I would expect the network overhead to also be several times smaller. Can you double check these results? }}
    \label{fig:ycsb_breakdown}
\end{figure}

Figure~\ref{fig:ycsb_breakdown} presents the runtime breakdown of \work \woundwait for workloads \texttt{A} and \texttt{C}.
We do not include results for Sherman/Deft since it uses memory server CPUs only for auxiliary operations.
Caching reduces indexing overhead by allowing memory server CPUs to directly locate data using hints from compute server caches, bypassing index traversal.
However, its overall impact is limited since network remains the primary bottleneck, as each transaction request requires a network round-trip.
Intra-transaction batching and inter-transaction batching effectively mitigate this bottleneck by significantly reducing the number of network round-trips.
While these optimizations enable more balanced utilization across system components, their impact is comparatively limited in workload \texttt{A} due to frequent read-write conflicts.

\begin{figure}[!t]
    \centering
    \includegraphics[width=0.47\textwidth]{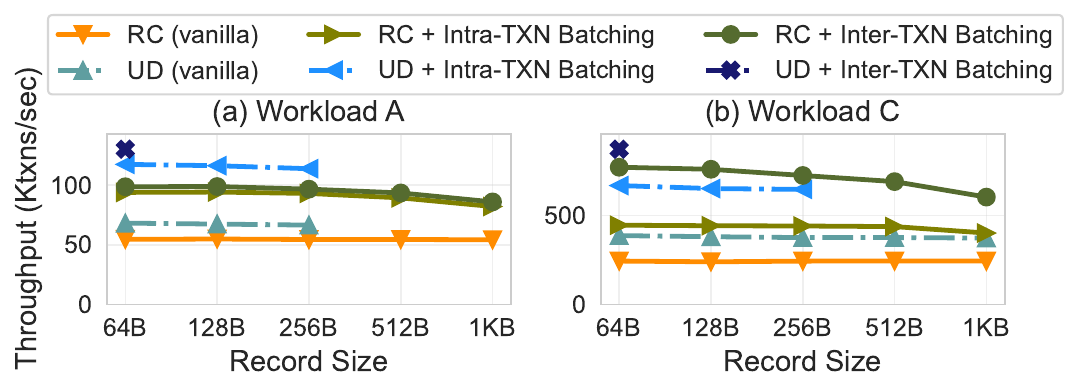}\vspace{-.15in}
    \caption{Throughput of Lotus on different RDMA transports under varying record sizes, in YCSB-\texttt{A} and \texttt{C}.}\vspace{-.15in}
    \label{fig:ycsb_row_size}
\end{figure}

\subsection{Analysis on Unreliable Datagram Transport}
\label{sec:rc_ud}

Next, we evaluate \work across different RDMA transports, including RC and UD, as prior studies on two-sided RDMA have primarily focused on UD~\cite{fasst, herd, eRPC}.
We decompose the batching factors of \work, i.e., intra-transaction batching and inter-transaction batching, and incrementally apply them to vanilla \work since UD transport has limited support for large message sizes, as discussed in Section~\ref{sec:batch_discussion}.

Figure~\ref{fig:ycsb_row_size} shows the throughput of \work \woundwait on RC and UD transports in workloads \texttt{A} and \texttt{C} under varying record sizes.
Overall, RC-based schemes generally achieve lower throughput than UD-based ones, due to additional overhead in reliability mechanisms such as acknowledgments and in-order delivery guarantees.
However, UD-based schemes fail to support large record sizes in both workloads, resulting in missing data points in the figure.
This limitation arises as network message sizes exceed the hardware MTU, as discussed in Section~\ref{sec:batch_discussion}.
Those large messages cannot be transmitted using UD without explicit application-level handling, which incurs additional network round-trips for packet fragmentation and reassembly of the out-of-order fragmented packets.
In contrast, RC supports those features at the transport layer, making it more flexible for workloads with varying record sizes.

While the results show several advantages of RC over UD, the choice of transports is orthogonal to \work’s design. 
\work is compatible with both RC and UD, as well as other transport protocols, and can benefit from improvements in the underlying transport.
% While the results highlight several advantages of RC over UD, the choice of transports is orthogonal to the design of \work. 
% \work is compatible with both RC and UD, as well as other transport protocols, and can potentially benefit from improvements in the underlying transport layer.
% \yxy{probably also worth pointing out that the choice between RC and UD and any other protocols that ppl might develop is orthogonal to the design of Lotus; Lotus is compatible with all of them and can also benefit if a better transport layer is used. We now argue that RC is better than UD in many aspects but in practice Lotus does not really care which is better. } \hcha{Added a short paragraph at the end.}

\subsection{Comparison with DEX}
\label{sec:dex_comparison}
Next, we compare \work with DEX~\cite{dex}, the state-of-the-art one-sided RDMA-based indexing scheme that employs compute-side partitioning to eliminate cache coherence across compute servers.
In this setting, each compute server logically owns a specific data partition so that cache coherence across multiple compute servers can be completely eliminated.
We focus on indexing performance in this experiment, since both \work and DEX can perform concurrency control locally in compute servers without remote memory accesses, as discussed in Section~\ref{sec:rdma_cc}.

\subsubsection{Scalability Analysis}
\begin{figure}[!t]
    \centering
    \includegraphics[width=0.47\textwidth]{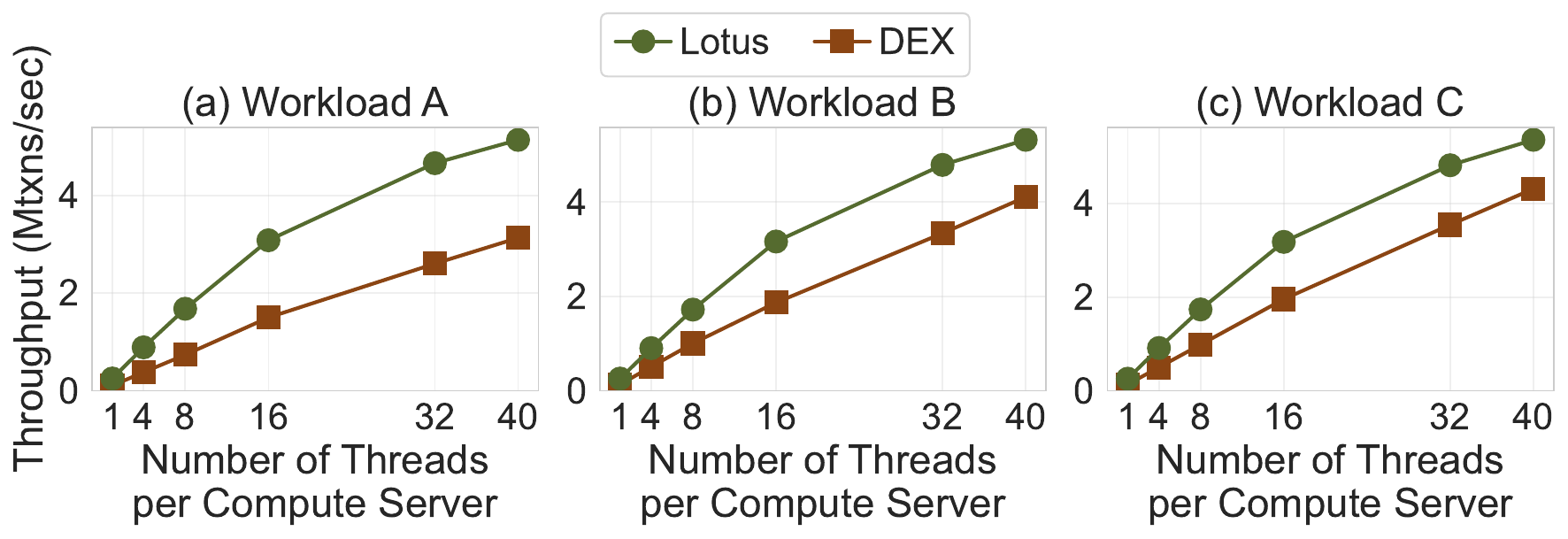}\vspace{-.15in}
    \caption{Throughput with a varying number of compute threads in YCSB workloads.}\vspace{-.15in}
    \label{fig:index_general_throughput}
\end{figure}

Figure~\ref{fig:index_general_throughput} shows the throughput of \work and DEX in YCSB workloads with a varying number of compute threads.
\work consistently outperforms DEX as it leverages a finer granularity of caching.
\work caches individual records that are frequently accessed, while DEX caches index nodes that may contain many infrequently accessed ones.
This allows \work to exploit better cache locality, especially in skewed workloads~\cite{twotree, anticaching, siberia}.

The performance gap between \work and DEX is larger in workload \texttt{A} than in \texttt{B} and \texttt{C}, due to the higher frequency of write operations that modify cache entries.
% This is because a higher number of write operations in workload \texttt{A} frequently modify cache entries.
% , because more write operations in workload \texttt{A} modify cached entries frequently.
Upon cache eviction of dirty entries, both \work and DEX must write back updated data to memory servers, incurring remote accesses.
% When evicting a modified cache entry, both \work and DEX must write the updated data back to memory servers, causing remote data accesses.
\work reduces this overhead by aggregating multiple write-back requests into a single network message, thereby minimizing the number of network round-trips.
However, DEX handles each write-back request in an either way: direct data update via RDMA WRITE or operation pushdown via RDMA SEND/RECV, incurring one network round-trip per request.

% \begin{table}[!t] 
%     \centering
%     \caption{Average network traffic per transaction under YCSB workloads.}
%     \label{tab:network_traffic}
%     \begin{tabular}{l | c | c | c}
%         \toprule[0.8pt]
%               & \textbf{Workload A} & \textbf{Workload B} & \textbf{Workload C} \\
%         \hline\hline
%         \work & 0.22~KB             & 0.21~KB             & 0.2~KB \\
%         DEX   & 12.09~KB            & 6.09~KB             & 5.65~KB \\
%         \bottomrule[0.8pt]
%     \end{tabular}
% \end{table}

\begin{figure}[!t]
    \centering
    \includegraphics[width=0.47\textwidth]{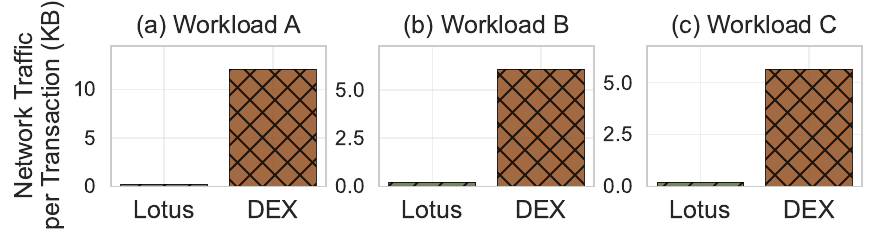}\vspace{-.15in}
    \caption{Network traffic in YCSB workloads.}\vspace{-.2in}
    \label{fig:index_network_traffic}
\end{figure}

\subsubsection{Network Traffic Analysis}
Now, we compare the network traffic of \work and DEX.
As in Section~\ref{sec:network_traffic}, for \work, we measure the traffic of two-sided RDMA operations.
For DEX, we measure both one-sided RDMA operations and two-sided RDMA operations for pushdown requests as it employs opportunistic offloading to leverage the benefits of both RDMA types.

Figure~\ref{fig:index_network_traffic} shows the average network traffic per transaction for each index under YCSB workloads.
Overall, the co-partitioned memory configuration incurs substantially less traffic than the non-partitioned memory setup analyzed in Section~\ref{sec:network_traffic}.
This reduction, particularly in workload \texttt{A}, stems from differences in memory models. Caching in co-partitioned memory avoids remote accesses, while non-partitioned memory systems still require remote accesses for cache coherence across compute servers, as discussed in Section~\ref{sec:disaggregation}.
Moreover, in the non-partitioned memory setting, most traffic is due to aborted transactions. In this experiment, we focus on indexing, which does not introduce transaction aborts.
 
\work generates significantly less network traffic than DEX across all workloads.
Upon a cache miss, \work only exchanges request information and the response, which are only few bytes in size, and further reduces the number of exchanges through batching.
% Upon a cache miss, \work only sends request information, e.g., operation type and index key, receives the response, e.g., value, and further reduces the number of network messages through batching.
In contrast, DEX needs multiple RDMA READs for index traversal upon a cache miss, and each node access incurs at least three RDMA READs to ensure correct synchronization~\cite{rdma_guideline}.
% Moreover, each node access requires at least three dedicated RDMA READs for correct synchronization~\cite{rdma_guideline}.
In particular, DEX generates much higher traffic than \work in workload \texttt{A}, a write-intensive workload.
Upon eviction of a modified cache entry, \work writes back only the updated record, whereas DEX writes back the entire index node containing that record.
Although DEX employs operation pushdown that can reduce data movement, it is an opportunistic optimization and may still trigger multiple network round-trips when child nodes reside on different memory servers.
% Note that the different configuration of the memory models accounts for the significant traffic gap between \work and DEX, compared to the smaller gap between \work and Sherman/Deft, since cache accesses in copartitioned memory do not directly trigger remote memory accesses to memory servers.

% \yxy{the traffic difference between Lotus and Sherman was not this significant. Explain why the difference is bigger here. What's different from the Sherman case.} \hcha{Added the second paragraph explaining the difference.}\yxy{In your explanation, can you also describe why the difference was not this big in non-partitioned memory scenario?}

\begin{figure}[!t]
    \centering
    \includegraphics[width=0.47\textwidth]{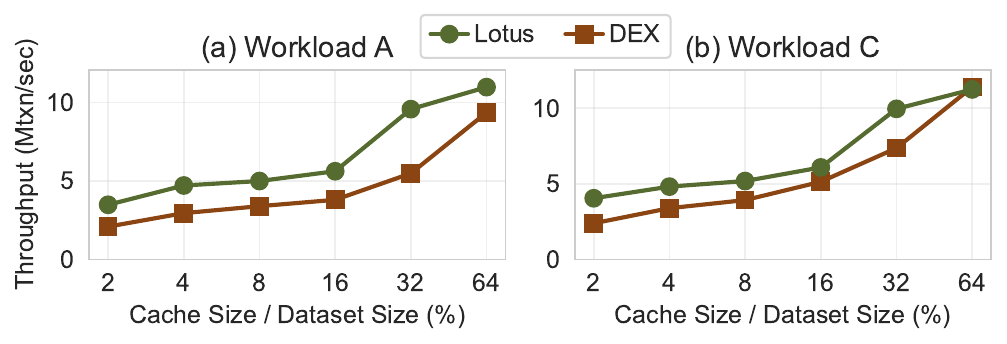}\vspace{-.17in}
    \caption{Throughput under different cache sizes in YCSB-\texttt{A} and \texttt{C}.}\vspace{-.18in}
    \label{fig:index_cache_throughput}
\end{figure}

\subsubsection{Cache Sensitivity Analysis}

Next, we study the impact of cache size on index performance.
Figure~\ref{fig:index_cache_throughput} shows the throughput of \work and DEX under varying cache sizes for workloads \texttt{A} and \texttt{C}.
% As cache size increases, both indexes benefit from local memory accesses.
In workload \texttt{A}, \work outperforms DEX due to its lightweight write-back mechanism.
In workload \texttt{C}, \work shows better performance with small cache sizes as its fine-grained record-level caching improves cache utilization.
As cache size grows, their performance eventually becomes comparable as both indexes can effectively serve most requests from local cache.
% Their performance eventually becomes comparable in large cache size settings, e.g., 64~\% of the dataset size, as both indexes can effectively serve most requests from local cache.

% \subsection{Impact of Memory Server CPUs}
\subsection{Sensitivity on Memory Server CPUs}
\label{sec:mem_threads_sensitivity}

\begin{figure}[!t]
    \centering
    \includegraphics[width=0.47\textwidth]{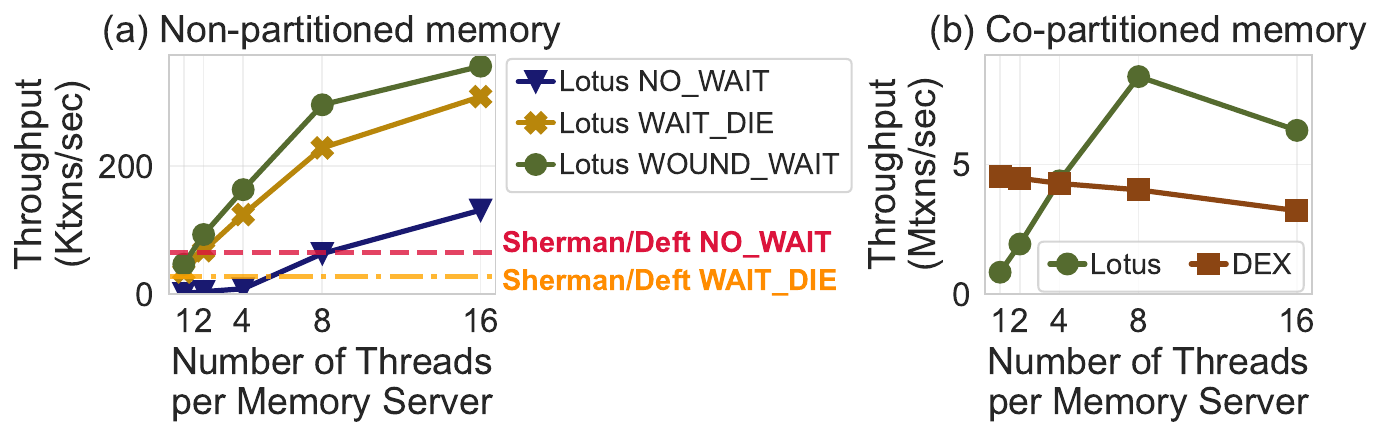}\vspace{-.15in}
    \caption{Throughput under a varying number of memory server threads in YCSB-\texttt{B}.}\vspace{-.15in}
    \label{fig:mem_cpu_sensitivity}
\end{figure}

Finally, we examine the performance impact of computing resources on memory servers under both non-partitioned memory and co-partitioned memory. 
We vary the number of memory server threads and adjust the number of compute threads accordingly.
% We fix the number of threads per compute server to 40 to maximize compute load and vary the number of threads per memory server.

Figure~\ref{fig:mem_cpu_sensitivity}(a) shows the performance under non-partitioned memory.
\work shows increasing throughput as the number of threads increases, effectively leveraging parallelism in both network and data processing.
We report the performance of Sherman/Deft with its best configuration since memory server threads are only used for auxiliary tasks such as remote memory allocation and registration.
% In contrast, Sherman/Deft maintains constant performance, since memory server threads are used only for auxiliary tasks such as remote memory allocation and registration.

Figure~\ref{fig:mem_cpu_sensitivity}(b) shows the performance under co-partitioned memory.
\work continues to scale up to 8 threads per memory server, but its performance decreases at 16 threads.
This is because memory servers become underutilized when fewer compute threads issue remote requests.
In particular, caching in co-partitioned memory fundamentally reduces the number of remote accesses.
DEX also utilizes memory server threads for operation pushdown.
However, reducing the number of compute threads has a more negative impact on performance than increasing memory server threads.

Overall, these results show that \work outperforms Sherman/Deft in non-partitioned memory and achieves comparable performance to DEX in co-partitioned memory, even under constrained resources.
They also indicate that memory server CPU allocation can serve as an effective performance tuning knob.

% \yxy{Figure 17 seems to suggest we should use more than 8 threads in the memory server? Since the performance is still increasing. If we use 16 memory server threads, does it mean we have to use fewer compute server threads?} \hcha{Yes, I believe the performance should increase further, but we would have to allocate fewer compute server threads for 16 memory server thread setting.}
\section{Related work}
\label{sec:related}

\noindent\textbf{RDMA indexes.}
Various index designs have been proposed to efficiently leverage RDMA.
% NAM-tree~\cite{namtree} explores three B+-tree variants based on one-sided RDMA, two-sided RDMA, and a hybrid of the two.
NAM-tree~\cite{namtree} explores three alternative B+-tree designs that can be implemented with one-sided RDMA, two-sided RDMA, and a hybrid of the two.
Sherman~\cite{sherman} focuses on one-sided RDMA and uses shortcut hint caching to reduce RDMA READs during index traversal.
Deft~\cite{deft} extends Sherman by segmenting index nodes to reduce the size of RDMA READs.
ROLEX~\cite{rolex} replaces B+-tree nodes with learned models that are much smaller in size, reducing network traffic during index traversal.
% ROLEX~\cite{rolex} leverages learned models to replace B+-tree nodes, which are much smaller in size, to reduce the network traffic in tree traversal.
DEX adopts a co-partitioned memory model to avoid remote access for cached data retrieval. 
% DEX~\cite{dex} adopts a co-partitioned memory model to avoid cache coherence across compute servers, reducing the traffic upon cache eviction.
% While the previous studies have tried to address the network amplification, \work does not suffer from the problem since only the request information and its result are transferred over the network with two-sided RDMA.
While these systems primarily focus on one-sided RDMA to mitigate network amplification, \work leverages two-sided RDMA to reduce network traffic at the source.
% While these systems primarily focus on one-sided RDMA to mitigate network amplification, \work handles it using two-sided RDMA to significantly reduce network traffic.
% \yxy{maybe less provocative if we say prior work focuses on 1-sided but Lotus focuses on 2-sided (is it true?)} \hcha{Yes, I edited the last sentence to be less provocative.}

\noindent\textbf{Shared memory databases with RDMA.}
Early studies on RDMA databases have explored various design alternatives for distributed transactions.
Pilaf~\cite{pilaf} adopts a hybrid approach, performing one-sided RDMA for read requests while handling write requests with two-sided RDMA.
FaRM~\cite{farm, farm2, farm_opactiy} exploits one-sided RDMA to minimize remote CPU involvement.
HERD~\cite{herd} and FaSST~\cite{fasst} discuss the drawbacks of one-sided RDMA, i.e., making system software inherently complex, by revisiting data processing with two-sided RDMA.
DrTMs~\cite{drtm, drtmh} leverage hardware transactional memory and RDMA to transform distributed transactions into local transactions.
% While the prior work has taught us valuable lessons, they target monolithic server architectures, which are not suitable for deployments in disaggregated architectures.
While prior work targets monolithic server architectures, we consider disaggregated architectures, which lead to different design focuses.
Our optimization techniques focus on reducing network round-trips to mitigate the overhead of data processing and network processing in memory servers.
% \yxy{this sentence is mildly criticizing prior work. Just objectively say we focus on different architectures, which leads to different design focuses (even better if we can be a bit more concrete here--why different arch leads to different design focuses?)} \hcha{I edited the last part.}

\noindent\textbf{Disaggregated memory databases with RDMA.}
NAM-DB~\cite{namdb} proposes a disaggregated memory database by leveraging one-sided RDMA.
RCC~\cite{rcc} evaluates various concurrency control protocols and provides insights to transform those legacy protocols to support one-sided RDMA.
In this work, we tackle the fundamental limitations of the one-sided RDMA approaches in indexing and concurrency control, 
i.e., network amplification and lack of functionality, 
and provide practical solutions using two-sided RDMA that support starvation prevention, priority-based scheduling, and preemption.
We also propose caching and batching techniques to further reduce network I/O while improving system performance.

% \noindent\textbf{SmartNIC systems.}
% SmartNIC, an emerging technology, has recently been used to accelerate systems by offloading host computations to to multi-core processors~\cite{liquidio, mips, bluefield2, bluefield3} and FPGAs~\cite{azure_smartnic, nexus, inova2, xlinx} in the NIC.
% There are active research efforts on SmartNIC integration in database applications such as key-value stores~\cite{xenic, kvdirect,smartnic_guideline}, storage systems~\cite{dds, dpdpu}, and data analytics~\cite{smartshuffle, dshuffle}.
% While our work does not directly target those systems, our optimization techniques can be also applied to the systems to further improve their performance.\yxy{maybe get rid of this part since we don't talk about SmartNIC in the paper.}

\noindent\textbf{CXL memory systems.}
Compute Express Link (CXL) memory is an emerging technology that enables memory sharing via PCI Express interconnects.
It is being actively explored for various system designs, including as remote caches~\cite{pond, directCXL}, key-value stores~\cite{bonsaiKV, chash}, and in-memory databases~\cite{saphanaCXL1, saphanaCXL2, polarCXLMem}.
While CXL memory has the potential to become a new standard for memory disaggregation, it is still in its early stages and limited to rack-scale deployments.
In this work, we focus on RDMA-based memory disaggregation, which is widely deployed in modern cloud environments.
\section{Conclusion}
\label{sec:conclusion}
This paper revisits the long-standing debate between one-sided RDMA and two-sided RDMA in the context of disaggregated memory databases.
We present \work, a two-sided RDMA-based disaggregated memory OLTP system, and challenge the prevailing view that one-sided RDMA is inherently superior due to limited computing resources in memory servers.
\work enables rich functionality for indexing and concurrency control and turns this limitation into a strength by leveraging caching and batching to minimize network round-trips and efficiently utilize remote CPU resources.
% \work addresses this limitation by leveraging caching and batching techniques to minimize network round-trips and efficiently utilize limited remote CPU resources.
Our performance study demonstrates that \work outperforms one-sided RDMA-based schemes across various workload scenarios.

% \begin{acks}
%This work was supported (in part) by the [...] Research Fund of [...] (Number [...]). Additional funding was provided by [...] and [...]. We also thank [...] for contributing [...].
% \end{acks}

\bibliographystyle{ACM-Reference-Format}
\bibliography{_bibliography.bib}

% \appendix

\end{document}